\documentclass[times,astrobib,amssymb]{mn2e}
\usepackage{psfig}

\newif\ifAMStwofonts

\newcommand{\lapp}{\mbox{\raisebox{-0.3em}{$\stackrel{\textstyle <}{\sim}$}}}
\newcommand{\gapp}{\mbox{\raisebox{-0.3em}{$\stackrel{\textstyle >}{\sim}$}}}
\def\kms{km~s$^{-1}$}

\title[21-cm absorption towards radio cores]{Associated 21-cm absorption towards the cores of radio galaxies}
\author[Yogesh Chandola, Neeraj Gupta and D.J. Saikia ]
       {Yogesh Chandola,${^1}$$\thanks{E-mail:chandola@ncra.tifr.res.in (YC), nrjgupta@gmail.com (NG), djs@ncra.tifr.res.in (DJS)}$  Neeraj Gupta$^{2}$ and D.J. Saikia$^{1,3}$   \\
$^1$ National Centre for Radio Astrophysics, TIFR, Pune 411 007, India \\
$^2$ Netherlands Institute for Radio Astronomy (ASTRON), Postbus 2, 7990 AA, Dwingeloo, The Netherlands \\
$^3$ Cotton College State University, Panbazar, Guwahati 781 001, Assam, India \\
}

\date{Accepted.    Received }

\pagerange{\pageref{firstpage}--\pageref{lastpage}}
\pubyear{  }

\begin{document}

\maketitle

\label{firstpage}

\begin{abstract}
We present the results of Giant Metrewave Radio Telescope (GMRT) observations to  
detect H{\sc i} in absorption towards the cores of a sample of radio galaxies.
From observations of a sample of 16 sources, we detect H{\sc i} in absorption
towards the core of only one source, the FR\,II radio galaxy 3C\,452 which has been 
reported earlier by Gupta \& Saikia (2006a). In this paper we present the results 
for the remaining sources which have been observed to a similar optical depth as 
for a comparison sample of compact steep-spectrum (CSS) and giga-hertz peaked 
spectrum (GPS) sources. We also compile available information on H{\sc i} 
absorption towards the cores of extended radio sources observed with angular 
resolutions of a few arcsec or better. The fraction of extended sources with
detection of H{\sc i} absorption towards their cores is significantly smaller (7/47)
than the fraction of H{\sc i} detection towards CSS and GPS objects (28/49). For the 
cores of extended sources, there is no evidence of a significant correlation 
between H{\sc i} column density towards the cores and the largest linear size  
of the sources.  The distribution of the relative velocity of the 
principal absorbing component towards the cores of extended sources is not   
significantly different from that of the CSS and GPS objects. However, a few
of the CSS and GPS objects have blue-shifted components $\gapp$1000 km s$^{-1}$,
possibly due to jet-cloud interactions. With the small number of detections
towards cores, the difference in detection rate between FR\,I (4/32) and FR\,II (3/15)
sources is within the statistical uncertainties.
\end{abstract}


\section{Introduction}
%
%
\begin{table*}
\caption{Radio sources observed with the GMRT.}
\begin{center}
\small{ 
 \begin{tabular}{ l l l l l l r l l l l }
\hline
Source & Alt. &  Redshift & S$_{\rm{1.4}}$ &  P$_{\rm{1.4}}$   & Type & LAS                         & LLS & LAS  & VLBI   \\
name   & name &           &  mJy                &  $\times$10$^{25}$    &      &                     &     & ref. & ref.   \\
       &      &           & (mJy)               &  (W\,Hz$^{-1}$)       &      & ($^{\prime\prime}$) &(kpc)&      &        \\
  ~~(1)& ~(2) &   ~(3)    &  ~(4)               &   ~(5)                & ~(6) & ~(7)                & (8) & (9)  &  (10)     \\
\hline
J0209+3547 &4C+35.03      &0.0377& 2087 & 0.70  &  FR\,I   &83        &61    & 0   & 1 \\
J0223+4259 &3C\,66B       &0.0213& 6217 & 0.65  &  FR\,I   &678       &288   & 1   & 2\\
J0313+4120 &S4            &0.1340&  487 & 2.38  &  FR\,II  &570       &1340  & 2   & 3\\
J0334+3921 &4C+39.12      &0.0206& 1183 & 0.12  &  FR\,I   &402       &165   & 1   & 2\\
J0418+3801 &3C\,111       &0.0485& 15121& 8.59  &  FR\,II  &195       &182   & 0   & 4\\
J0656+4237 &4C42.22       &0.0590&  937 & 0.80  &  FR\,I   & 67       &76    & 0   & -\\                                                          
J0748+5548 &DA240         &0.0357& 2237 & 0.67  &  FR\,II  &2040      &1428  & 3   & 5\\                                                          
J0758+3747 &3C\,189       &0.0428& 2637 & 1.16  &  FR\,I   &144       &120   & 0   & 2\\
J1147+3501 &              &0.0631& 815  & 0.80  &  FR\,I   &780       &936   & 4   & 2\\                                                          
J1628+3933 &3C338         &0.0304& 3595 & 0.78  &  FR\,I   &112       &67    & 0   & 2\\                                                          
J1744+5542 &NGC6454       &0.0312& 810  & 0.18  &  FR\,I   &200       &123   & 5   & 3\\                                                          
J1835+3241 &3C\,382       &0.0579& 5548 & 4.55  &  FR\,II  &171       &189   & 6   & 2\\
J1836+1942 &PKS           &0.0161& 1458 & 0.09  &  FR\,I   &316       &102   & 1   & -\\
J1842+7946 &3C390.3       &0.0555&11537 & 8.73  &  FR\,II  &235       &250   & 0   & 6 \\                                                          
J2245+3941&3C\,452        &0.0811&10383 & 17.25 &  FR\,II  &256       &386   & 7   & 2\\
J2338+2701 &3C465         &0.0302& 7590 & 1.63  &  FR\,I   &487       &290   & 1   & 7\\
\hline
\end{tabular}
}
\end{center}
\begin{flushleft}
Column 1: source name; column 2: alternative name;
column 3: redshift ;
column 4: flux density at 1.4 GHz in mJy;
column 5: luminosity at 1.4 GHz; 
column 6: radio morphology i.e. FR\,I or FR\,II;  
columns 7 and 8: largest projected angular (LAS) and linear (LLS) size in arcsec and kpc respectively, where LAS
has been estimated from the outer hotspots in the FR\,II sources, and from the outermost contours in the FR\,I sources;  
column 9: references for the LAS;
column 10: references for the VLBI structure.\\
References for the largest angular sizes -- 0: This paper;  1: NRAO VLA Sky Survey (NVSS; Condon et al. 1998);
2: de Bruyn (1989); 3: Willis, Strom \& Wilson (1974); 4: Giovannini et al. (1999); 5: Bridle \& Fomalont (1978); 6: Leahy \& Perley (1991);  
7: Gupta \& Saikia (2006a) \\  

All red-shift information  is from NED (NASA Extragalactic Database) except for J1744+552 and J1842+7946. 
The redshift for J1744+552 has been taken from Bridle \& Fomalont (1978) and for J1842+7946 has been taken from Eracleous \& Halpern (2004). \\

 Flux densities at 1.4 GHz are determined using the maps from NVSS ; Luminosities have been calculated
from these flux densities assuming a spectral index of 1.\\

 References for the VLBI structure: 1: Lara et al. (1997); 2: Giovannini et al. (2001); 3: Britzen et al. (2007); 4: Linfield (1987); 5: Saripalli et al. (1997); 6: Alef et al. (1996); 7: Venturi et al. (1995)\\

\end{flushleft}

\label{cores_sample}
\end{table*}

It is widely believed that the source of energy of active galactic nuclei (AGN) 
is the accretion of matter onto a supermassive black hole with an accretion disk,
located in the centre of the galaxy. While there have been many theoretical and 
numerical studies of accretion flows (e.g. Hawley 2011),
observational studies of the gaseous components in the circumnuclear regions
could provide useful insights towards understanding the dynamics and properties
of the gas that might be fuelling the AGN. At radio frequencies an important method
of probing this region is via 21-cm H{\sc i} absorption towards compact radio 
sources, such as the cores of radio galaxies and quasars, and the compact 
steep-spectrum (CSS) and gigahertz peaked-spectrum (GPS) sources. The CSS and GPS 
objects (O' Dea 1998) typically have sizes of $\lapp$10 and 1 kpc respectively, 
and are believed to be the young ($<$10$^5$ yr) progenitors of the larger radio galaxies and quasars 
which extend up to a few Mpc, with the largest sources being typically $\sim$10$^8$ 
yr old (Jamrozy et al. 2008; Konar et al. 2008). Such studies could also help us study the evolution of the gaseous
component with the age of the radio source and also test consistency of these
properties with the unified schemes for AGN (Pihlstr\"om, Conway \& Vermeulen 2003; Gupta \& Saikia 2006b;  
Curran \& Whiting 2010). 

Studies of H{\sc i} absorption towards radio galaxies and quasars have shown
that the absorption lines are more often detected towards CSS and GPS objects, with previous
studies (Pihlstr\"om, Conway \& Vermeulen 2003; Gupta et al. 2006; Chandola, Sirothia \& Saikia 2011) showing 
an apparent anti-correlation between the integrated optical depth and source size. For an assumed
fixed value of the covering fraction and spin temperature this corresponds to an anti-correlation with the
neutral hydrogen column density. 
Gupta \& Saikia (2006b) found the 21-cm absorption detection rate to be higher towards the 
CSS and GPS sources identified with galaxies than with quasars.  
They find that there is a tendency for the detection rate as well as the column density for galaxies to increase 
with core prominence, a statistical indicator of the orientation of the jet axis to the line of sight. This can be 
understood in a scenario where radio sources are larger than the scale of the circumnuclear H{\sc i} disk so that the lines of 
sight to the lobes at very large inclinations do not intersect the disc.
On the other hand, 
Curran \& Whiting (2010) and more recently Allison et al. (2012) have investigated the effect of UV luminosity on the 
21-cm absorption detection rate.  They found that after excluding the sources with UV luminosities, $L_{UV}$$\gapp$10$^{23}$\,W\,Hz$^{-1}$, 
the 21-cm absorption detection rate for compact radio sources is only marginally higher than for other radio sources. 
Curran \& Whiting suggest that higher 21-cm absorption detection rate amongst compact objects is probably 
due to their generally low UV luminosity rather than their compact sizes or orientation.

Compared with the studies of H{\sc i} absorption 
towards CSS and GPS objects, the extended radio sources have received less attention. However, there 
have been a few notable exceptions like van Gorkom et al. (1989), Morganti et al. (2001), and more recently Emonts et al. (2010).  
To understand any evolution in the H{\sc i} component of the circumnuclear gas as the radio source grows from the GPS and 
CSS scales to larger dimensions, it is necessary to probe the gas on similar 
scales from radio observations probing the compact regions with comparable optical depths. 
This can be achieved via 21-cm H{\sc i} observations towards the radio cores 
of large radio sources with resolutions of about a few arcsec or better so that
the core is resolved from the more extended emission and can be unambiguously identified. 
Such observations will also enable us to determine any differences in the H{\sc i} properties of FR\,I
and FR\,II sources. There have been suggestions that the mode of fuelling the 
nucleus and the structure of the torus/disk may be different for the FR\,I and 
FR\,II radio sources, with those in FR\,Is being geometrically thin (e.g. Morganti
et al. 2001). 

In this paper we present the results of 21-cm H{\sc i} absorption observations
towards the cores of FR\,I and FR\,II radio galaxies with the Giant Metrewave Radio
Telescope (GMRT), and combine our results with those in the literature to 
examine any evolution in the H{\sc i} properties in the circumnuclear region
with source size. Possible differences between FR\,I and FR\,II radio galaxies,
and the distribution of the velocities of the absorbing components relative to
the systemic velocities have also been examined. 
The observations are described in Section~\ref{cores_obs}. The
discovery of 21-cm absorption towards the core of the FR\,II radio galaxy 
J2245+3941 (3C\,452), which is a part of our sample, has been reported earlier 
(Gupta \& Saikia 2006a). In this paper we present the results for the remaining 
fifteen sources. In Section \ref{cores_disc} we combine our 
results with those of similar searches and discuss the statistical trends in 
gas properties with respect to radio-source characteristics. The results are 
summarized in Section \ref{cores_sum}.


\section{Observations} 
\label{cores_obs}
%
%
\begin{table*}
\caption{Observational details and results of the search for associated H{\sc i} absorption.}
\begin{center}
\vspace{5mm}
 \begin{tabular}{ l l r c c r l c c c c }
\hline
Source&~~Date&Obs. freq& BW &t$_{\rm{int}}$&  Beam~~~   &Comp. &Peak flux & $\sigma$ & $\tau$ & $N$(H{\sc i})\\
      &    & (MHz)    &(MHz)&  (hrs)&($^{\prime\prime}\times^{\prime\prime}$ $^\circ$) & & (mJy/b)  &  (mJy/b)  & (10$^{-3})$ & (10$^{20}$cm$^{-2}$)\\
~~(1)   & ~~~(2)  &  (3)~~~   & (4)  &  (5)  &   (6)~~~~~    &  (7)   &  (8)   &   (9)   &    (10) & (11)  \\
\hline
J0209+3547 & 2006 Jan & 1368.76& 4 & 3.5 & 2.41$\times$1.91 ~+20  &Core&102   & 0.91    &  $<$26.7  & $<$5.15 \\
J0223+4259 & 2006 Jan & 1390.84& 4 & 4.6 & 2.64$\times$2.25 ~$-$83&Core&170   & 1.10    &  $<$19.5  & $<$3.77 \\
J0313+4120 & 2004 Jan & 1252.56& 8 & 3.0 & 2.62$\times$2.15 ~$-$53&Core&383   & 1.12    &  $<$8.7  & $<$1.68 \\
J0334+3921 & 2006 Jan & 1391.75& 4 & 4.9 & 2.70$\times$1.99 ~+53  &Core&304   & 0.85    &  $<$8.4  & $<$1.62 \\
J0418+3801 & 2006 Jan & 1354.70& 4 & 5.2 & 3.15$\times$2.38 ~+34  &EHS &1393  & 0.90    &  $<$2.1  & $<$0.41 \\
           &          &        &   &     &                        &Core&1103  & 1.05    &  $<$3.0  & $<$0.58 \\
           &          &        &   &     &                        &WHS &250   & 1.03    &  $<$12.3  & $<$2.37 \\
J0656+4237 & 2010 Jan & 1341.27& 4 & 4.9 & 3.63$\times$2.80 ~+64  &Core&224  & 0.53     &  $<$6.9  & $<$2.95 \\
J0748+5548 & 2010 Jan & 1371.49& 4 & 4.4 & 2.70$\times$2.16 ~+25  &Core&209   & 0.38    &  $<$5.4  & $<$2.78 \\
J0758+3747 & 2006 Jan & 1362.06& 4 & 5.2 & 3.12$\times$2.07 ~+43  &Core&184   & 1.01    &  $<$16.5  & $<$3.19 \\
J1147+3501 & 2010 Jan & 1336.06& 4 & 4.4 & 3.69$\times$2.38 ~+38  &Core&351   & 0.57    &  $<$4.8  & $<$1.96 \\
J1628+3933 & 2010 Jan & 1378.56& 4 & 4.8 & 3.42$\times$2.17 ~+61  &EP  &67.6  & 0.93    &  $<$41.4 & $<$7.99 \\
           &          &        &   &     &                        &Core&179   & 1.14    &  $<$19.2  & $<$3.70 \\
           &          &        &   &     &                        &WP  &69.5  & 0.76    &  $<$32.7 & $<$6.32 \\
J1744+5542 & 2010 Jan & 1377.43& 4 & 4.0 & 4.24$\times$3.18 ~$-$75  &Core&313 & 1.18    &  $<$11.4  & $<$2.20 \\
J1835+3241 & 2006 Mar & 1342.70& 4 & 4.5 & 2.68$\times$2.10 ~+35  &EHS &203   & 1.05    &  $<$15.6  & $<$3.01 \\
           &          &        &   &     &                        &Core&240   & 1.17    &  $<$14.7  & $<$2.84 \\
           &          &        &   &     &                        &WHS &37    & 1.12    &  $<$90 & $<$17.4 \\
J1836+1942 & 2004 Jan & 1397.84& 4 & 2.0 & 3.04$\times$2.10 ~$-$70&Core&224   & 2.10    &  $<$28.2  & $<$5.44 \\
J1842+7946 & 2010 Jan & 1345.72& 4 & 4.9 & 3.88$\times$2.32 ~+17   &NHS &177   & 1.02    &  $<$17.4  & $<$3.34 \\
           &          &        &   &     &                        &Core&165   & 1.15    &  $<$18.9  & $<$4.02 \\
           &          &        &   &     &                        &SHS &839   & 1.90    &  $<$6.9  & $<$1.31 \\
J2245+3941$^{*}$ & 2005 Dec & 1313.85& 4 & 5.6 & 3.14$\times$2.25 ~+52&EHS &94& 1.10    &  $<$35.1 & $<$6.77 \\
           &          &        &   &     &                        &Core&194   & 1.00    &   58.0  & 6.39 $\pm 0.13$ \\
           &          &        &   &     &                        &WHS &128   & 1.10    &  $<$25.8  & $<$4.98 \\
J2338+2701 & 2004 Jan & 1378.74& 4 & 4.0 & 2.89$\times$2.24 ~$-$50&Core&302   & 1.63    &  $<$16.2  & $<$3.13 \\
\hline
\end{tabular}
\end{center}
\begin{flushleft}
Column 1: source name; column 2: year and month of observations; column 3:  redshifted 21-cm frequency;
column 4:  baseband bandwidth (BW);
column 5:  observing time in hr (excluding
calibration overheads); columns 6, 7 and 8: restoring beam, radio component (EHS = eastern hotspot, 
WHS = western hotspot, NHS = northern hotspot, SHS = southern hotspot, EP = eastern peak, WP = western peak and Core= radio core) and peak
brightness in mJy/beam for the continuum image made using
line-free channels respectively; column 9: rms noise in the spectrum; column 10:
optical depth estimate, and column 11:  H{\sc i} column density in units of 
10$^{20}$ cm$^{-2}$, assuming ${T}_{\rm s}$=100 K and a covering factor, $f_c$ of unity; upper limits are 3$\sigma$ values,
assuming $\Delta v$=100 \kms.\\
$^{*}$ The 21-cm absorption towards J2245+3941 (3C\,452) has velocity of $-$2.8 km s$^{-1}$ relative to its optical systemic velocity.\\
\end{flushleft}
\label{cores_GMRTobs+res}
\end{table*}

%
Table~\ref{cores_sample} lists the 16 FR\,I and FR\,II radio galaxies 
which we observed with the GMRT over the period of 2004$-$2010. Table~\ref{cores_sample}
also lists some of their observed properties. With the available observing
time, these 16 were selected largely from the 3CR and B2 samples to have a core flux density at 1400
MHz $\gapp$100 mJy with resolutions of a few arcsec or better so that the core
could be distinguished from the more extended bridge/jet emission.  The 16
sources observed with the GMRT consist of 10 FR\,I and 6 FR\,II radio sources.  
 Their projected linear sizes range from $\sim$ 61 kpc to as large as $\sim$ 1.4 Mpc 
($H_0=$71 km s$^{-1}$ Mpc$^{-1}$, $\Omega_m=$0.27, $\Omega_{vac}=$0.73; Spergel et al. 2003).

The observations were made with a baseband bandwidth (BB BW) of 4 MHz (velocity range $\sim$ 800 km s$^{-1}$) split into 
128 channels (resolution $\sim$ 7 km~s$^{-1}$) , except for J0313+4120 for which a BB BW of 8 MHz
(resolution $\sim$ 15 km~s$^{-1}$) was used 
(see Table~\ref{cores_GMRTobs+res}). For the observations of J2245+3941 (3C\,452) on 
2005 Dec 11 the then new high-resolution mode (resolution $\sim$ 3.6 km~s$^{-1}$) of the GMRT correlator was used 
(cf. Gupta \& Saikia 2006a). The expected redshifted 21-cm frequency was placed
at the centre of the BB BW by tuning the local oscillator chain. 
The standard flux density and bandpass calibrators (3C\,48, 3C\,147 and 3C\,286) 
were usually observed every three hours to correct for variations in amplitude and
bandpass. Observations of the target source were preceded and followed by observations
of a phase calibrator, a compact radio source, typically every 35 min. 

The Astronomical Image Processing System ({\tt AIPS}) package was used to reduce the
data. After flagging bad data, including those affected by Radio Frequency Interference 
(RFI), and calibrating the data, a  continuum image of the source was made by averaging
$\sim$60 line-free channels. The images were made with the highest resolution, which
was typically $\sim$2 arcsec with the GMRT at L-band, because of our interest in 
detecting absorption towards the compact components such as the radio cores. The data
were self-calibrated to produce a satisfactory continuum image; after which the 
complex gains were applied to all the frequency channels and the continuum image was
subtracted from the visibility data cube. Spectra were extracted at the pixels of
maximum brightness in the cores and the hot-spots. This was done independently for
both Stokes RR and LL to check for consistency. The two polarization channels were 
later combined to produce the final Stokes I spectrum which was shifted to the 
heliocentric frame.   The results for all the 16 sources, including  
the radio galaxy J2245+3941 (3C\,452) are summarised in Table~\ref{cores_GMRTobs+res}. 

\begin{table*}
\caption{Sources included in `cores sample' from literature}
\begin{center}
 \begin{tabular}{ l l l l l l l l l l r r l }
\hline
Source & Alt. & Redshift&  S$_{\rm{1.4}}$ & P$_{\rm{1.4}}$             & LAS                 &  LLS   & LAS & VLBI& Type  & $N$(H{\sc i}) & V$_{shift}$ & Ref. \\
name   & name &         &                     & $\times$10$^{25}$      &                     &        & ref. & ref.&       & $\times$10$^{20}$ & & \\
        &     &         &   (mJy)             &  (W\,Hz$^{-1}$)        & ($^{\prime\prime}$) & (kpc)  &      &     &       &  (cm$^{-2}$) &  (km s$^{-1}$) & \\
(1)    & (2)  &  (3)    &       (4)           & (5)                    &    (6)              &  (7)   & (8)  & (9) & (10)   & (11) & (12) &  (13)   \\
\hline
J0037$-$0109    & 3C15      & 0.0730 &  4092    & 5.46 & 33.7        & 46     &1    & -  & FR\,II           &$<$0.50 &      & M \\
J0055+3021      & NGC315    & 0.0165 &  2371    & 0.15 & 3480        & 1152   &2    & 1 & FR\,I             &  7.0   &  462 & E\\
J0057$-$0123    & 3C29      & 0.0450 &  5354    & 2.60 & 139         & 121    &1    & -  & FR\,I            &$<$9.39 &      & M \\
J0107+3224      & 3C31      & 0.0169 &  4757    & 0.31 &  900        & 305    &3    & 1 & FRI             & $<$0.6 &      & E \\          
J0126$-$0120    & 3C40      & 0.0180 &  3879    &0.29  & 339         & 122    &1    & -  & FR\,I            &$<$4.50 &      & M \\
J0308+0406      & 3C78      & 0.0287 &  7360    &1.42  & 210         & 119    &4    & 2  & FR\,I            &$<$0.73 &      & M \\
J0329+3947      &  -        & 0.0243 &  1419    &0.20  & 386         & 186    &5    & 3  & FRI             &$<$2.8  &      & E \\
J0331+0233      & 3C88      & 0.0302 &  5030    &1.08  & 191         & 114    &1    & -  & FR\,II           &$<$5.06 &      & M \\
J0429$-$5349    &   -       & 0.0399 &  5830$^{\ast}$  &2.21  & 276  & 215    &6    & -  & FR\,I            &$<$11.46&      & M \\
J0519$-$4546    & PicA      & 0.0351 & 66000$^{\ast}$  &19.24 & 458  & 316    &7    & 4  & FR\,II           &$<$2.43 &      & M \\
J0621$-$5241    & MRC       & 0.0511 &  3400$^{\ast}$  &2.15  & 47.7 & 47     &1    & 5  & FR\,I            &$<$3.60 &      & M \\
J0927+2955      &  -        & 0.0253 &  297     & 0.04 & 747         & 376    &8    & -  & FRI             &$<$69   &      & E \\
J1006+3453      & 3C236     & 0.1005 &  4645    & 12.07& 2307        & 4221   &9    & 6  & FR\,II           &   3.7  &  162 & G \\
J1111+2657      & NGC3563   & 0.0331 &  99      & 0.03 &  72         & 47     &10   & -  & FRI             &$<$3.7  &      & E \\
J1145+1936      & 3C264     & 0.0217 &  5494    &0.60  & 540         & 234    &11   & 7  & FR\,I            &$<$1.54 &      & G \\
J1230+1223      & M87       & 0.0044 & 153420   &0.67  & 896         & 81     &12   & 8,9 & FR\,I           &$<$1.35 &      & G \\
J1254$-$1233    & 3C278     & 0.0150 &  7622    &0.39  & 128         & 38.7   &13   & 10,11 & FR\,I           &$<$12.32&      & M \\
J1321$-$4342    & NGC5090   & 0.0114 &  5860$^{\ast}$& 0.17 & 770    & 177    &9    & 5 & FR\,I            & 2.70   &   4  & M \\
J1323+3133      & NGC5127   & 0.0162 &  741     & 0.04 & 688         & 223.6  &10   & 1  & FRI             & 4.9    &  33  & E \\ 
J1324+3622      & NGC5141   & 0.0174 &  886     & 0.06 &  52         & 18.15 &14   & -  & FRI             & 3.0    &  134 & E \\
J1336$-$3356    & IC4296    & 0.0125 & 11960$^{\ast}$& 0.43 & 1869   & 470.9  &1    & 5  & FR\,I           &$<$7.43 &      & M \\
J1516+0701      & 3C317     & 0.0345 &  5608    & 1.57 & 41.4        & 28.0   &1    & 5  & FR\,I           &$<$1.74 &      & G \\
J1632+8232      & NGC6251   & 0.0247 &  2201    & 0.31 & 3150        & 1547   &15   & 12,13& FR\,I         &$<$2.51 &      & G \\
J1644$-$7715    & MRC       & 0.0427 &  6500$^{\ast}$& 2.79 & 259.8  & 215.9  &1    &  - & FR\,II          &$<$8.63 &      & M \\
J1720$-$0058    & 3C353     & 0.0304 &  54835   &11.84 & 252.9       & 151.7  &1    & 12 & FR\,II          &   42   &$-$100& M \\
J1806+6949      & 3C371     & 0.0510 &  2138    & 1.35 & 280         & 275    &16   & 14 & FR\,I           &$<$2.90 &      & G \\
J1952+0230      & 3C403     & 0.0590 &  5855    & 5.00 &  98.5       & 111    &17   &  - & FR\,II          &$<$17.82&      & M \\
J2118+2626      & NGC7052   & 0.0156 &  212     & 0.01 & 149         & 46.6   &18   &  - & FRI             &$<$2.4  &      & E \\ 
J2157$-$6941    & PKS       & 0.0283 &  30390$^{\ast}$ & 5.66 &78.8  & 47.3   &19   & 15 & FR\,II          &$<$1.90 &      & M \\
J2223$-$0206    & 3C445     & 0.0559 &  5405    & 4.10 & 570.0       & 611.0  &20   &  - & FR\,II          &$<$8.06 &      & M \\
J2231+3921      & 3C449     & 0.0171 &  2847    & 0.19 &  1500       & 514.5  &3    &  - & FRI             &$<$1.2  &      & E \\
\hline
\end{tabular}
\end{center}
\begin{flushleft}
%
Column 1: source name; column 2: alternative name; column 3: redshift; column 4: 1.4GHz flux density  
column 5: 1.4GHz luminosity in the rest frame of the source;
columns 6 \& 7: largest projected angular (LAS) and linear (LLS) size
in arcsec and kpc respectively; column 8: references for the LAS;
column 9: references for the VLBI structure;
column 10:  radio morphology classification; column 11: H{\sc i} column density or a 3$\sigma$ upper limit to it;
column 12: the shift of the primary H{\sc i} component relative to the systemic velocity as measured from the optical
emission lines, with a negative sign indicating a blue-shift, and column 13: references for the H{\sc i} observations.\\
References for the LAS: 1:Morganti et al. (1993); 2: Bridle et al. (1976); 3: Andernach et al. (1992);  4: Saikia et al. (1986); 5: Bridle et al. (1991); 6: Jones \& McAdam (1992); 7: Perley, Roser \& Meisenheimer (1997); 8: Ekers et al. (1981);  9: Mack et al. (1997);  10: Parma et al. (1986);  11: Gavazzi, Perola \& Jaffe (1981); 12: Owen, Eilek \& Kassim (2000); 13: Baum et al. (1988); 14: Fanti et al. (1986); 15: Perley, Bridle \& Willis (1984); 16: Wrobel \& Lind (1990); 17: Dennett-Thorpe et al. (1999); 18: Morganti et al. (1987); 19: Fosbury et al. (1998); 20: Leahy et al. (1997)  \\

References for H{\sc i} observations: M: Morganti et al. (2001); G: van Gorkom et al. (1989); E: Emonts et al. (2010)\\
  Flux densities at 1.4 GHz are determined using maps from NVSS except the sources with $^{\ast}$,  values for which are from 
Wall \& Peacock (1985); Luminosities  have been calculated from these flux densities assuming spectral index of 1.\\
 References for the VLBI structure: 1: Giovannini et al. (2001) ; 2: Jones (1984); 3: Giovannini et al. (2005); 4: Tingay et al. (2000); 5: Venturi et al. (2000); 6: Schilizzi et al. (2001); 7: Lara et al. (1997); 8: Chang et al. (2010); 9: Kovalev et al. (2007); 10: Marecki et al. (2003); 11: Bondi et al. (2004); 12: Fomalont et al. (2000); 13: Jones et al. (1986); 14: Gabuzda et al. (2004); 15: Ojha et al. (2004) \\
\end{flushleft}
\label{cores_fsample}
\end{table*}

%
\begin{table*}
\caption{Sources reported in literature for cores with H{\sc i} detection but not included in our  `cores sample'}
\begin{center}
 \begin{tabular}{ l l l l l l l l r r l }
\hline
Source & Alt. & Red-shift&  LAS            &  LLS   & LAS & Type  & $N$(H{\sc i}) & V$_{shift}$ & Ref. \\
name   & name &      &                     &        & Ref.&     & $\times$10$^{20}$ & & \\
        &     &      &    ($^{\prime\prime}$) &  (kpc)   &      &    &    (cm$^{-2}$)  &   (km\,s$^{-1}$)  & \\
(1)    & (2)  & (3)  &      (4)             & (5)    & (6) & (7) & (8) & (9) & (10)    \\
\hline

J0319+4130& NGC1275  & 0.0176 & 198         &  70    &1   & FR\,I &  2.34  & 2797 & YRS \\
J0840+2949& 4C29.30  & 0.0647 & 520         & 639    &2   & FR\,I &  47    &  84  & CSG \\
J0918$-$1205& Hydra A& 0.0549 & 310         & 326    &3   & FR\,I &  42    &$-$180& D \\
J1219+0549& NGC4261  & 0.0075 & 492         & 75     &4   & FR\,I &  7.2   &$-$3  & J \\
J1247+6723&          & 0.1073 & 671         & 1301   &5   & FR\,II&  6.73  & 57 & S \\
J1325$-$4301& Cen A  & 0.0018 & 30564       & 563    &6   & FR\,I &  50.1  & 6  & VGH \\
J1352+3126& 3C293    & 0.0450 & 216         & 189    &7   & FR\,I &  78.9  &  3 &B \\
J1531+2404& 3C321    & 0.0961 & 286         & 503    &8   & FR\,II&  92.3  &  235 & CSDN\\ 
J1959+4044& Cygnus A & 0.0561 & 127         & 137    &9   & FR\,II&  25.4  &  165 & C \\

\hline
\end{tabular}
\end{center}
\begin{flushleft}%
Column 1: source name; column 2: alternative name; 
column 3: redshift; columns 4 \& 5: largest projected angular (LAS) and linear (LLS) size
in arcsec and kpc respectively; column 6 : references for the LAS;
column 7: H{\sc i} column density;
column 8: radio morphology classification;
column 8: the shift of the primary H{\sc i} component relative to the systemic velocity as measured from the optical
emission lines, with a negative sign indicating a blue-shift, and column 10: references for the H{\sc i} observations.\\
References for the structural information and LLS: 1: Pedlar et al. (1990); 
2: Chandola, Saikia \& Gupta (2010) 3: Ekers \& Simkin (1983); 4: Condon \& Broderick (1988); 
5: Lara et al. 2001 ; 6: Junkes et al. (1993); 7: Bridle, Fomalont \& Cornwell (1981); 8: Chandola et al. (2012); 9: Perley, Dreher \& Cowan (1984).\\
References for H{\sc i} observations: D: Dwarakanath, Owen \& van Gorkom (1995); J: Jaffe \& McNamara (1994); 
S: Saikia, Gupta \& Konar (2007); C: Conway \& Blanco (1995); YRS: De Young, Roberts \& Saslaw (1973); VGH:  van der Hulst, Golisch \& Haschik (1983); CSDN: Chandola et al. (2012); CSG: Chandola, Saikia \& Gupta (2010); B: Beswick, Pedlar \& Holloway (2002)\\
%
\end{flushleft}
\label{cores_Dsample}
\end{table*}

\begin{figure}
\centerline{\vbox{
\psfig{figure=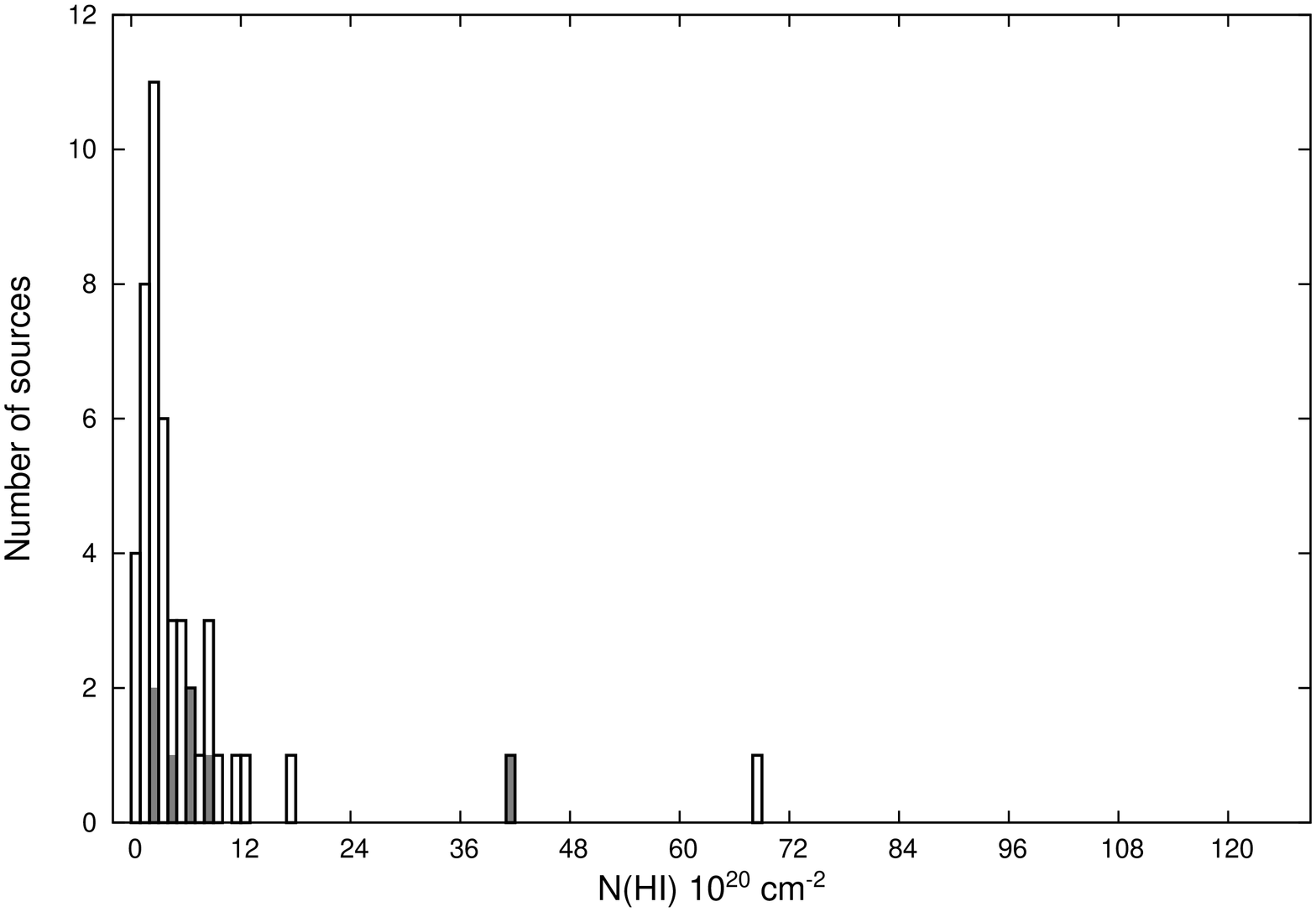,width=8.5cm,angle=-90} }}
\caption[Histogram showing H{\sc i} column density distribution of `cores sample']
{H{\sc i} column density distribution of the `cores sample'. Shaded areas represent the detections while the unshaded areas represent 3$\sigma$ upper limits.}
\label{cores_distribution} 
\end{figure}
\begin{figure}
\centerline{\vbox{
\psfig{figure=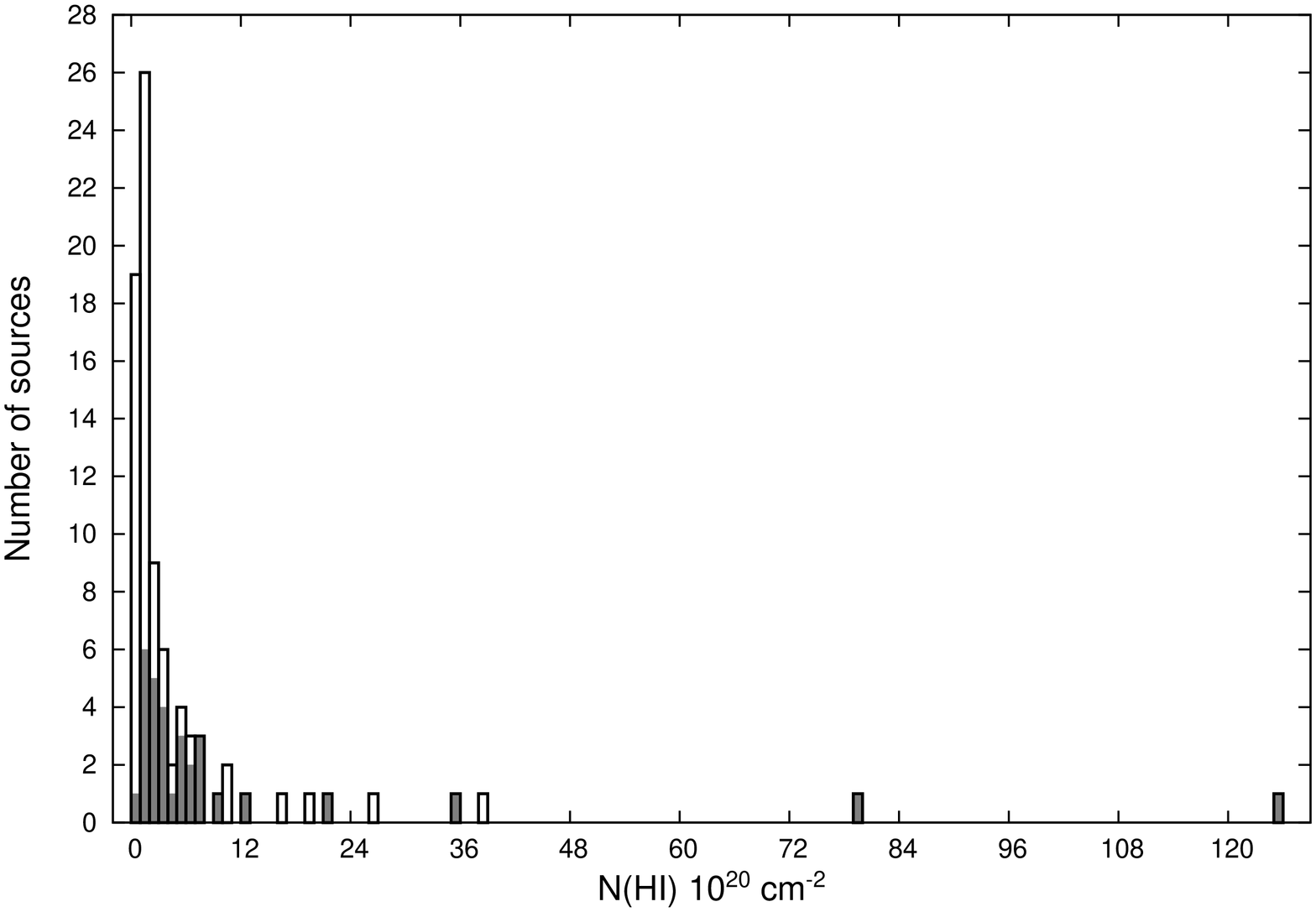,width=8.5cm,angle=-90} 
\psfig{figure=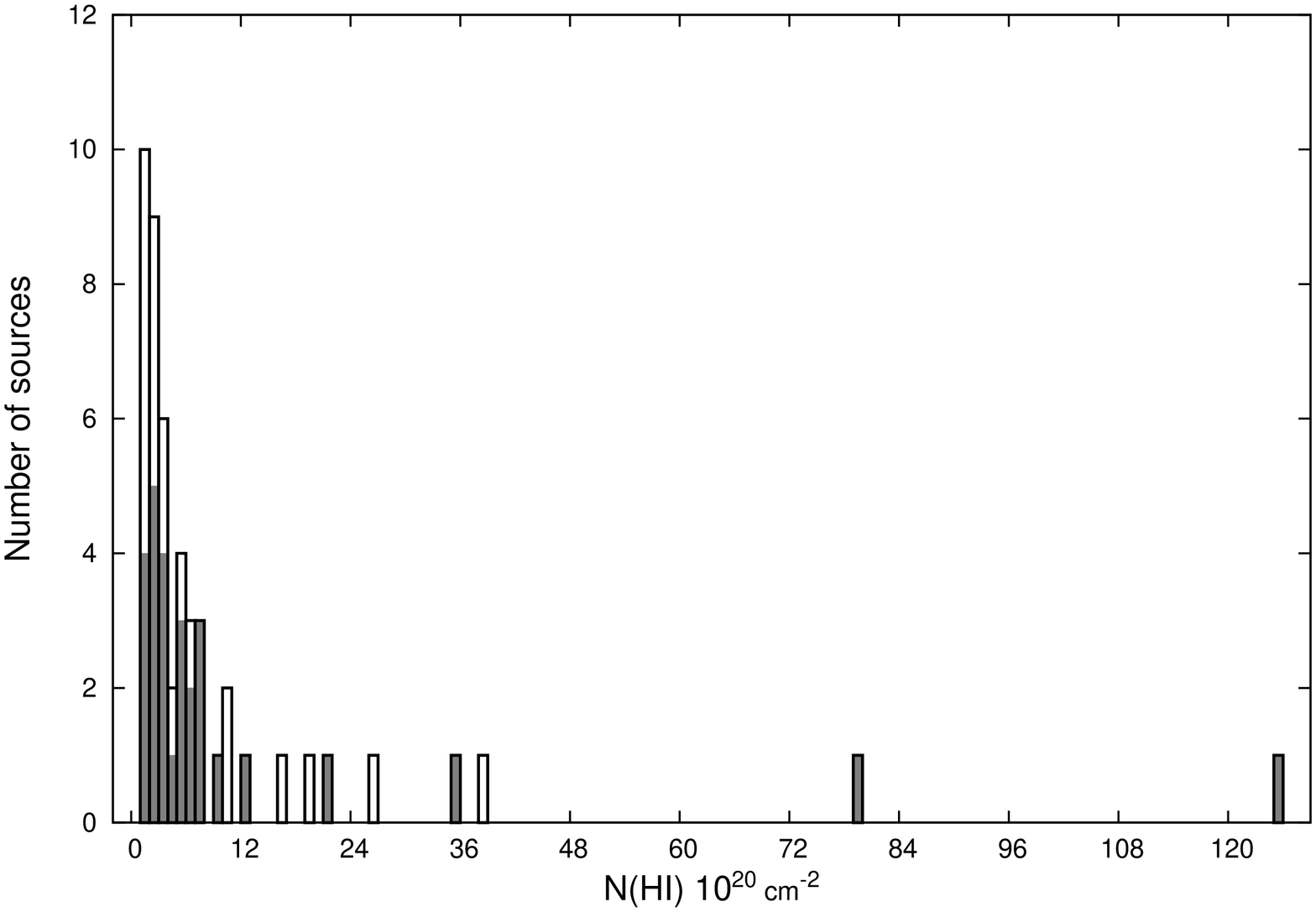,width=8.5cm,angle=-90} }}
\caption[Histogram showing H{\sc i} column density distribution of `CSS \& GPS sources']
{Upper panel: H{\sc i} column density distribution of the `CSS \& GPS sources' from Gupta et al. (2006). 
Lower panel: H{\sc i} column density distribution of CSS \& GPS sources with $N$(H {\sc i}) or upper limits to it greater 
than 1.5$\times$10$^{20}$ cm$^{-2}$. Shaded areas represent the detections while the unshaded areas represent 3$\sigma$ upper limits.}
\label{CSS_distribution} 
\end{figure}
%
%
\begin{figure}
\centerline{\vbox{
\psfig{figure=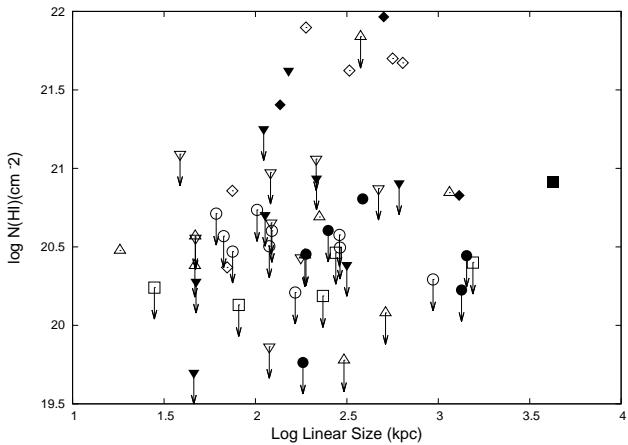,width=8.5cm,angle=-90} }}
\caption[H{\sc i} column density as a function of projected linear source size]{H{\sc i} column density as a 
function of projected linear source size.  Arrows mark the 3$\sigma$ upper limit to the column density.  
The unfilled symbols are FR\,I sources while filled symbols are FR\,II sources. Circles denote the sources from our GMRT observations while squares, 
triangles with vertex up and triangles with vertex down denote sources from Emonts et al. (2010), van Gorkom et al. (1989) and Morganti et al. (2001) 
respectively. Rhombuses are sources which are not included in our cores sample but have detection towards the cores of
larger radio sources with $\sim$ arcsec resolution.     
}
\label{cores_nhivssize} 
\end{figure}

\begin{figure}
\centerline{\vbox{
\psfig{figure=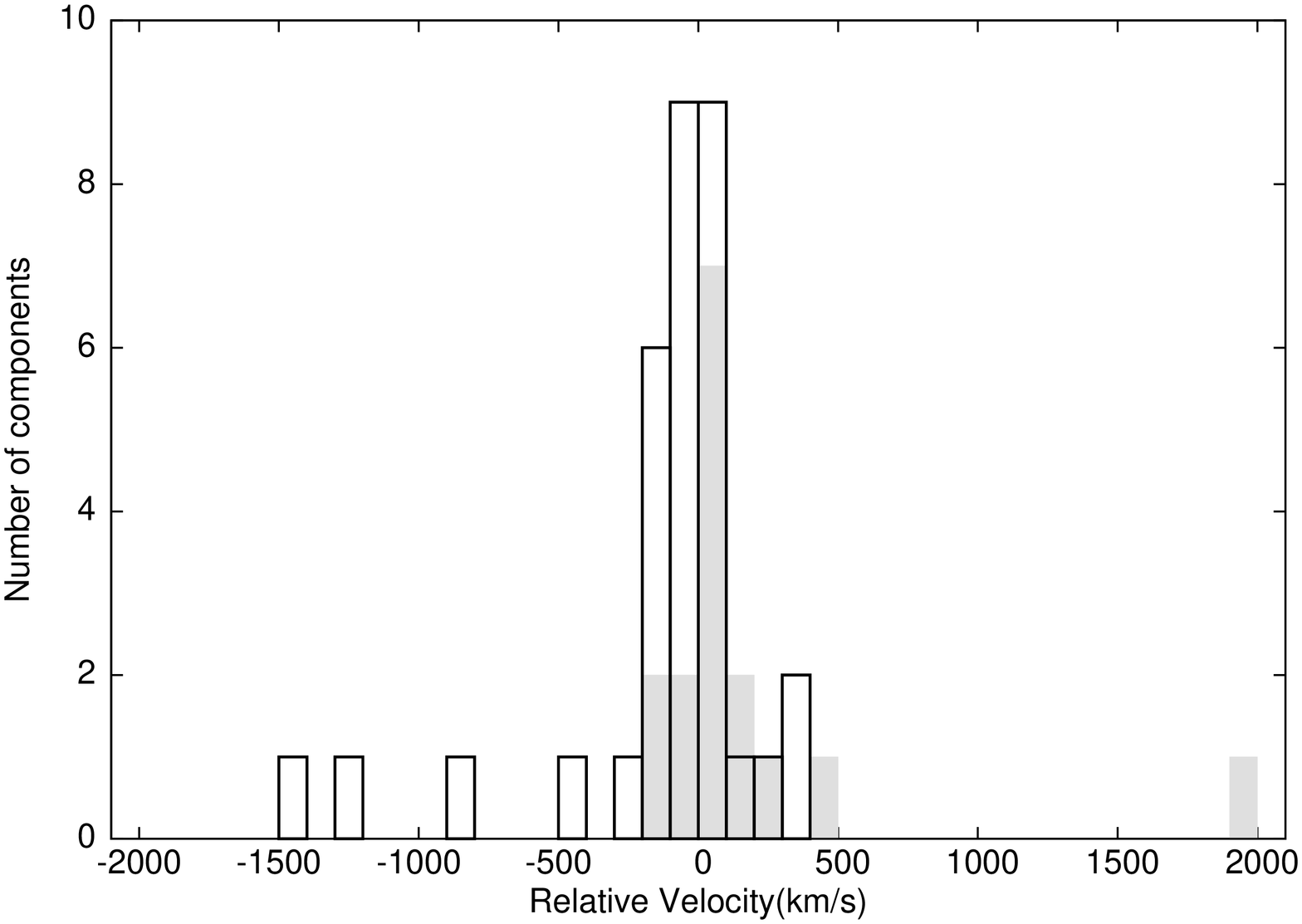,width= 8.5cm,angle=-90 }}}
\caption[Distribution of velocities for the principal H{\sc i} absorption components]{The distribution of
velocities for the principal H{\sc i} absorption components with respect to the systemic velocity estimated 
from optical lines. Ones with bold boxes represent the distribution for CSS \& GPS sources, while the shaded areas denote 
the cores of FR\,I and FR\,II radio sources. In the cores, Perseus A (NGC1275) where the 
principal relative velocity 2797 km/s, has been put in bin at 2000 km/s.} 
\label{cores_velhist}
\end{figure} 

The final radio continuum images, i.e. the ones used for continuum subtraction, and the 21-cm H{\sc i} absorption
spectra for the 15 galaxies with non-detections are presented in the Appendix 
(Figs.~\ref{j0206}$-$\ref{j2338}). The large-scale
structure is seen in most of the images except for 
J0313+4120, J0748+5548, J1147+3501, J1744+5542 and J1836+1942 where the diffuse emission is largely resolved out 
in our GMRT images. For all sources the  rms noise in the spectra, the column density for the lone detection, 3C452, and 
3-$\sigma$ upper limits to the optical depth towards all the other cores, and also towards the hotspots or peaks in 5 of the 
16 sources are presented in Table.~\ref{cores_GMRTobs+res}. The H{\sc i} column density, $N$(H{\sc i}), has 
been estimated using the relation

\begin{center}
\begin{tabular}{c c l r}

{$N$(H{\sc i})} &=& $1.823\times10^{18}$ {\Large $\frac{{{T}_{\rm s}}~\int{\tau(v)dv}}{f_c}$}~${\rm cm^{-2}}$ & ~~~~~(1) \\
\end{tabular}
\end{center}

and assuming  ${T}_{\rm s}$=100 K and $f_c$=1.0, where ${T}_{\rm s}$, $\tau$ and $f_c$ are the spin temperature, optical depth at a velocity
$v$ and the fraction of background emission covered by the absorber respectively.  In this paper wherever we mention H{\sc i} 
column density, it refers to the pseudo H{\sc i} column density with the above assumptions. The upper limits
to $N$(H{\sc i}) are 3-$\sigma$ values and have been estimated, as stated above, assuming ${T}_{\rm s}$=100 K, $\Delta v$=100 km s$^{-1}$
and $f_c$=1.0.

\section{Discussion}
\label{cores_disc}
In order to estimate reliably the fraction of sources with 21-cm H{\sc i} detection, it is important
to consider surveys of sources which report the detections as well as the non-detections. In our sample
of 16 radio galaxies observed with the GMRT we detect 21-cm absorption 
associated with the core of only 1 radio galaxy i.e. J2245+3941 (3C\,452) (Gupta \& Saikia 2006a).
To improve upon the statistics, we have considered the observations of large ($>$15 kpc) sources by van Gorkom et al. 
(1989), Morganti et al. (2001)  and Emonts et al. (2010) who have reported both the detections and non-detections in their samples,
and have summarized their results in Table~\ref{cores_fsample} which has a total of 31 sources.  We have included only 
those sources which have been observed with few arcsec resolution (corresponding to a beam size of $\lapp$15 kpc for our red-shift range) 
in order to minimise the effect of large scale structures on our estimates. We have also avoided inclusion of 
QSOs, BL Lacs, Spirals or Seyferts in the sample as we wanted to
have a sample of radio galaxies with their hosts being ellipticals. 
Thus combining our observations of 16 sources listed in Table~\ref{cores_sample} and the 31 sources
listed in Table~\ref{cores_fsample}, we obtain a sample of 47 radio sources, called the `cores sample'.  
We also list  9 sources in Table~\ref{cores_Dsample} for which H{\sc i} 21 cm absorption has been reported 
towards their core from observations with angular resolution of a few arcsec but not 
included in our `cores sample'  for statistical analysis since in these cases non-detections were not published.      


\subsection{Frequency of occurrence}
A comparison of the results of H{\sc i} 21-cm absorption observations towards the cores of radio galaxies 
with those obtained for CSS and GPS objects could help explore possible evolution of the content of H{\sc i}
gas in the host galaxies of radio sources as the sources evolve. The column densities or upper limits to these 
in our `cores sample' range from $\sim$$0.5\times$10$^{20}$ cm$^{-2}$ to  $69\times$10$^{20}$ cm$^{-2}$. The median column density
 sensitivity of our `cores sample', including the upper limits is $3.1\times$10$^{20}$ cm$^{-2}$.  
For comparison, we consider the compilation of CSS and GPS sources from Gupta et al. (2006).  From this compilation 
of 62 CSS and GPS sources we consider 60 sources.  We exclude J1415+1320 and J1945+7055 as these were reported as 
individual detections.  To enlarge this sample further we also consider 21-cm absorption searches based on samples of 
steep-spectrum sources reported in the literature since 2006. We consider measurements of 17 sources from 
Chandola, Sirothia \& Saikia (2011), 5 compact sources (B0258+35, B0648+27, B1122+39, B1447+27, B1557+26) with 
steep spectrum from Emonts et al. (2010), and 2 sources from Carilli et al. (2007).   
The measurements of 10 compact sources with steep spectrum are available from Allison et al. (2012) 
but detailed size information is not available for these.  So we are not able to include these in our sample.  
Thus we have a sample of 84 CSS and GPS sources. This sample has column densities or upper 
limits to these ranging from $\sim$$0.06\times$10$^{20}$ cm$^{-2}$ to $125\times$10$^{20}$ cm$^{-2}$ with a median value of 
 $1.9\times$10$^{20}$cm$^{-2}$ assuming that the values for the non-detections are close to the upper limits. 
To ensure that the sources are observed to similar sensitivity limits so that the distributions of column densities 
are similar for both the `cores' and the `CSS and GPS' samples, we considered all the sources from  `CSS and GPS sample'
with a column density estimate or an upper limit to it which is at least 1.5$\times$10$^{20}$ cm$^{-2}$. 
This leaves us with a sample of 49 CSS and GPS sources with a median column density sensitivity of $3.8$$\times$10$^{20}$ cm$^{-2}$, 
assuming again that the values for the non-detections are close to the upper limits.

The detection rate for the `cores sample' is rather low (7/47; $\sim$15\%;  see Fig.~\ref{cores_distribution} ) 
compared with the detection rate for compact CSS and GPS sources (28/49; $\sim$57\%;  see Fig.~\ref{CSS_distribution} lower panel).
 Considering the entire sample of 84  CSS and GPS objects, the detection rate (31/84; $\sim$37\%; 
see Fig.~\ref{CSS_distribution} upper panel ) is again significantly higher than for the `cores sample'. All but one 
of the sources from `CSS and GPS sample' with column density values $\leq$1.5$\times$10$^{20}$ cm$^{-2}$ represent sensitive
 observations of CSS and GPS objects with no detection of H{\sc i} in absorption. We have also examined the 
VLBI-scale structure (references given in Table~\ref{cores_sample} and Table~\ref{cores_fsample}) of the 
cores of the larger sources (`cores sample') and find that they tend to have a core-jet structure where 
most of the flux density is from the VLBI-scale core component which is usually smaller than the CSS and GPS
sources suggesting that the difference in  H{\sc i} absorption detection rate is not
due to different covering factors. Although it would be useful to confirm it from a larger sample,  
clearly the detection rate of H{\sc i} in absorption towards the cores of larger sources is smaller than for
the CSS and GPS objects, suggesting an evolution in the gaseous content of the host galaxies .

It is also interesting to examine the H{\sc i} detection rate for FR\,I and FR\,II radio sources in case this
reflects any differences in either the torus/disk or fuelling processes (e.g. Morganti et al. 2001;
Emonts et al. 2010). In the `cores sample' H{\sc i} is detected in absorption towards 4 out of 
32 ($\sim$13\%) FR\,I objects, compared with  3 out of 15 ($\sim$20\%) for the FR\,II sources. Within 
statistical errors, there is no significant difference in the detection rate towards the cores for 
FR\,I and FR\,II radio sources. These have been observed with similar sensitivity.

\subsection {Column density vs size}
As mentioned earlier, the H{\sc i} column density in CSS and GPS objects appears to 
be anti-correlated with source size ( Pihlstr\"om, Conway \&
Vermeulen 2003; Gupta et al. 2006; Chandola, Sirothia \& Saikia 2011), but shows no evidence 
of any dependence on either redshift or luminosity (Gupta et al. 2006).
On sub-galactic scales, the observed anti-correlation has been interpreted by Pihlstr\"om, Conway \&
Vermeulen (2003) to be due to gas distributions with a radial power law density profile. It is also
relevant to note here that Emonts et al. (2010) observed a sample of radio galaxies to detect H{\sc i}
in emission and found an inverse relation between H{\sc i} mass and sources size.

In Fig.~\ref{cores_nhivssize} we plot the 21-cm H{\sc i} column density towards the cores against the 
projected linear size in kpc for the sample of 47 radio galaxies. In this figure, we also plot an 
additional nine sources where H{\sc i} detection has been reported from observations with angular 
resolutions of a few arcsec for single sources (see Table~\ref{cores_Dsample}), giving a  total of
16 detections. The linear sizes range from tens of kpc to over a Mpc while the column densities vary 
from  0.5$\times$ 10$^{20}$ cm$^{-2}$ to 69$\times$10$^{20}$ cm$^{-2}$. There does not seem to be any
significant relation between the H{\sc i} column density towards the cores and the largest projected 
linear sizes of the sources.


\subsection{Relative velocity of absorbing gas}

The relative velocity of the absorbing gas could be significantly blue shifted if it
is due to gas which has been accelerated by interaction with the radio jet, while
infalling material fuelling the central engine would appear red shifted. An 
important component of the unification scheme for FR\,II radio galaxies is the
presence of a circumnuclear disk/torus. H{\sc i} gas associated with this and
rotating around the nucleus may appear both blue and red shifted relative to the
systemic velocity. It is worth noting that CSS and GPS objects often show complex
line profiles with both blue- and red-shifted H{\sc i} absorption components.

We compare the relative velocity of the principal absorbing component for the
16 detections towards cores of larger sources with the  33 detections(including 2 
sources from Gupta et al. (2006) which we left out earlier while doing statistical 
analysis of detection rates) towards CSS and GPS objects. While making these comparisons
it is also relevant to note that different ionization lines may yield somewhat
different systemic velocities. For example in 3C\,452, the low-ionization forbidden
emission lines are broader and more red-shifted than the high-ionization ones.
The typical error in optical systemic velocity estimate for the sample is $\sim$ 90 km~s$^{-1}$.
Fig.~\ref{cores_velhist} shows that although a few CSS and GPS objects have 
large blue-shifted velocities of $\gapp$1000 km s$^{-1}$ possibly due to jet-cloud
interactions, the distributions for the cores and CSS and GPS objects are not 
significantly different. Also one of the objects with the highly blue-shifted component, J1815+6127, 
has a large error of $\sim$900 km~s$^{-1}$ (Vermeulen \& Taylor 1995) in its optical systemic velocity. 
The highly red-shifted component seen towards the core of NGC1275 in the Perseus cluster could be 
due to a gas cloud or galaxy in the intracluster medium moving towards NGC1275 (De Young, Roberts \& Saslaw 1973).

%
\section {Summary}
\label{cores_sum}
We have presented the results from a search for 21-cm absorption towards the cores of 
16 FR\,I and FR\,II radio galaxies. From our search we have reported earlier one new 
detection, i.e. towards the core of the radio galaxy J2245+3941 (3C\,452) (Gupta \& Saikia 2006a),
and have now presented the upper limits on N(H{\sc i}) for the remaining 15 sources.  
We have combined our results with those of similar 21-cm absorption searches in the 
literature to obtain a larger sample, called the `cores sample', of 47 radio galaxies,
and have summarized our results here.
  
\begin{enumerate}
\item
The upper limits to the H{\sc i} column density towards the cores of large sources 
presented here range from $\sim$0.5 to 69$\times$10$^{20}$ cm$^{-2}$.

\item 
Considering observations of similar sensitivity, we find that the detection rate of
H{\sc i} absorption towards the cores of larger ($>$15 kpc) sources is only $\sim$15 per cent, 
compared with $\sim$57 per cent for CSS and GPS objects, suggesting an evolution in
the gaseous content of the host galaxies as the radio sources age and grow in size.

\item
The H{\sc i} column density towards the core vs the largest linear size plot for the 
large sources shows no significant correlation between these parameters. 
 
\item
The distribution of the relative velocity of the
principal absorbing component towards the cores of large sources is not
significantly different from that of the CSS and GPS objects, although a few 
of the CSS and GPS objects show large blue-shifted velocites ($\gapp$1000 km s$^{-1}$)
possibly due to jet-cloud interactions. 

\item
For the `cores sample', H{\sc i} absorption 
is detected towards 4 out of 32 ($\sim$13\%) FR\,I objects, and 3 out of 15 ($\sim$20\%) 
FR\,II sources. Although there have been suggestions of differences in the torus/disk structure
and fuelling mechanisms between FR\,I and FR\,II sources, with the small number of
detections there does not appear to be a 
significant difference in the detection rate between these two types of radio sources.

\end{enumerate}
%
\section*{Acknowledgments}
YC thanks Akash Pirya for his useful suggestions during this work.  
We thank the reviewer for his/her detailed comments which helped improve the manuscript, and
the GMRT staff for their help with the observations. The GMRT is a national facility
operated by the National Centre for Radio Astrophysics of the Tata Institute of Fundamental Research.
We also thank the numerous contributors to the GNU/Linux group. This research has made use of 
the NASA/IPAC Extragalactic Database (NED) which is operated by the Jet Propulsion Laboratory, 
California Institute of Technology, under contract with the National Aeronautics and Space 
Administration.  



\clearpage

\appendix

\section{Maps and 21-cm spectra  of the `cores sample' from our GMRT observations.}
\begin{figure}
\centerline{\vbox{
\psfig{figure=0206_map.ps,width=6.5cm,angle=-90}
\vskip -1.5 cm
\hskip 0.2 cm
\psfig{figure=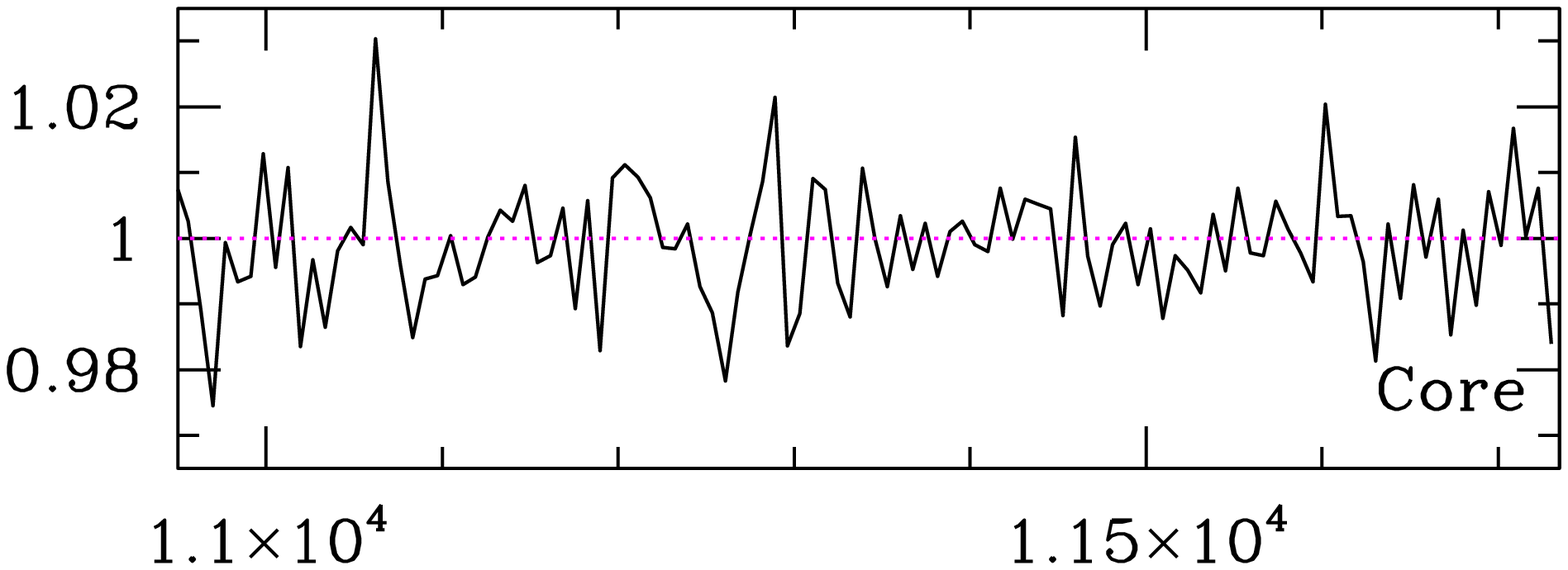,width=6.5cm,height=4cm,angle=0}
}}
\hskip +0.0cm
\vskip -1.5 cm
\caption[GMRT image and 21-cm absorption spectra of J0209+3547]
{Image of J0209+3547 (4C+35.03) with an rms noise of 0.2 mJy/beam.  The contour levels are
0.6$\times$($-1$, 1, 2, 4, 8, 16, 32, 64, 128) mJy/beam. In all the images the restoring beam is shown as
an ellipse while the position of the optical host galaxy is marked with a (+) sign.
The dashed contours represent the negative flux density values while solid contours represent the
positive flux density values.
The spectra at the peak intensity pixel (shown as ($\times$) sign) for the different components are shown 
below each image and the components have been indicated in the spectra.  For all 
the spectra the x- and y-axes represent the heliocentric velocity in km s$^{-1}$ and 
the normalised intensity respectively. The spectra have been normalised using the flux density 
corresponding to the peak intensity pixel.
}
\label{j0206}
\end{figure}
%

%
%

\begin{figure}
\centerline{\vbox{
\psfig{figure=0220_map.ps,width=6.5 cm,angle=-90}
\vskip -1.2cm
\hskip 0.2 cm
\psfig{figure=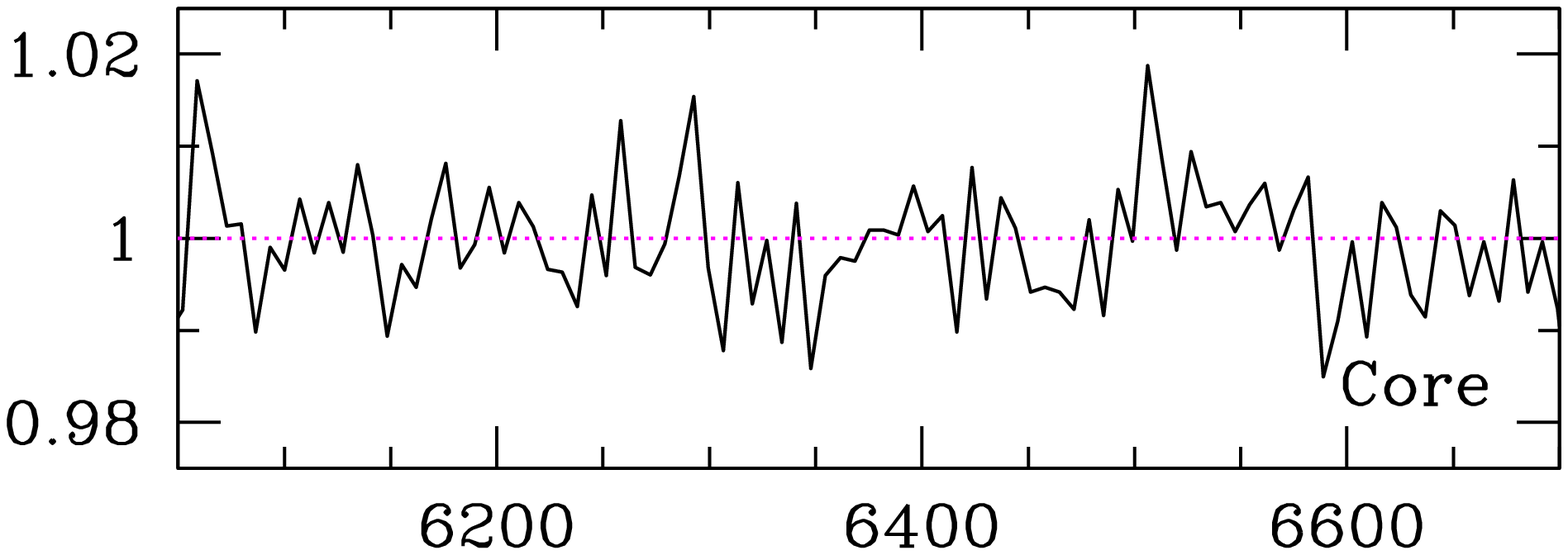,width=6.5cm,height=4cm,angle=0}
\vskip -1.5cm
}}
\caption[GMRT image and 21-cm absorption spectra of J0223+4259]{ Image of J0223+4259 (3C\,66B) with an rms noise of 0.6 mJy/beam.  The contour levels are
3$\times$($-1$, 1, 2, 4, 8, 16, 32, 45) mJy/beam.  
}

\label{J0220}
\end{figure}

%

\begin{figure}
\centerline{\vbox{
\psfig{figure=0309_MAP.PS,width=6.5cm,angle= 0}
\vskip -1.5cm
\psfig{figure=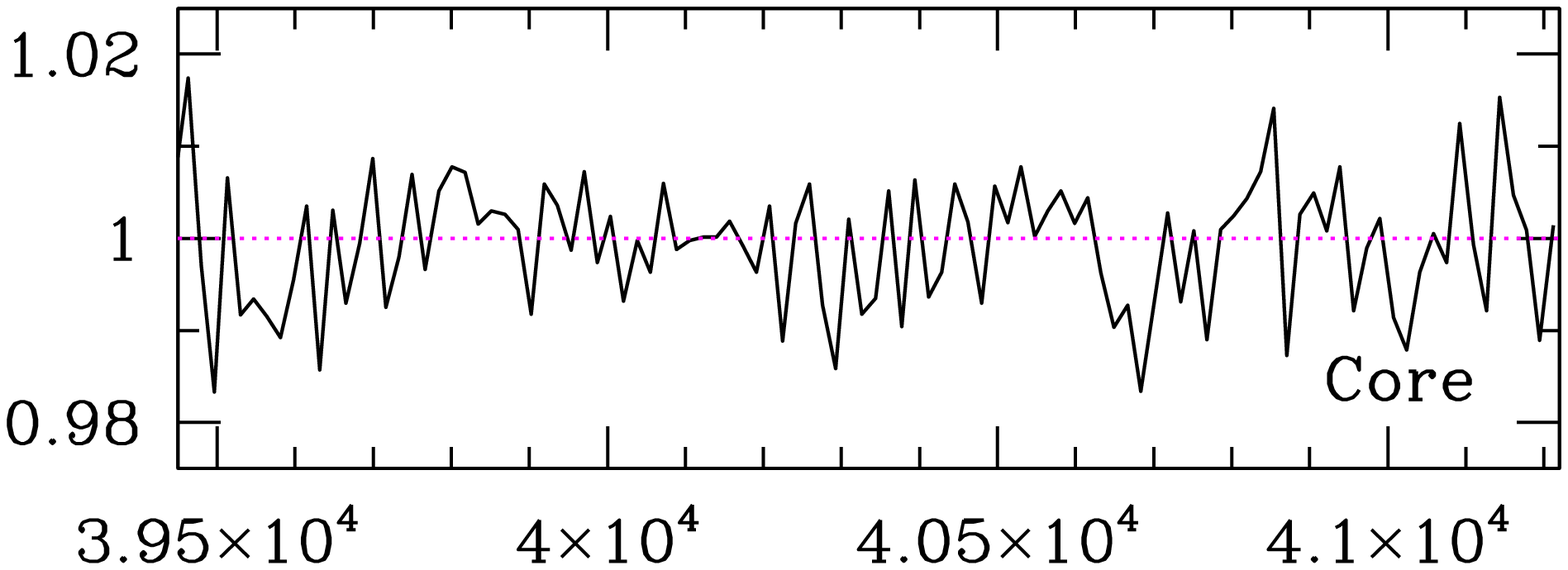,width=6.5cm,height=4cm,angle=0}
\vskip -1.5cm
}}
\caption[GMRT image and 21-cm absorption spectra of J0313+4120]
{Image of J0313+4120 with an rms noise of 0.5 mJy/beam.  The contour levels are
2.5$\times$($-1$, 1, 2, 4, 8, 16, 32, 64, 128) mJy/beam.  
}

\label{J0309}
\end{figure}

%
%

%
\begin{figure}
\centerline{\vbox{
\psfig{figure=0331_map.ps,width=6.5cm,angle=0}
\vskip -1.5cm
\psfig{figure=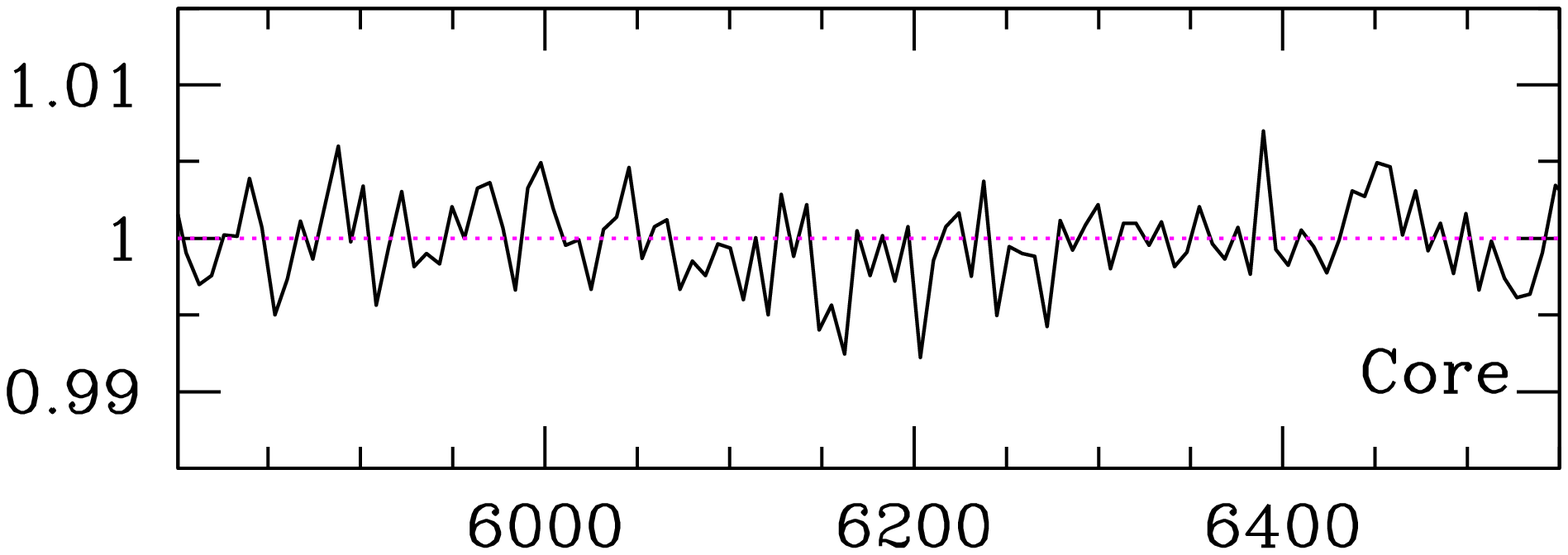,width=6.5cm,height=4cm,angle=0}
}}
\vskip -1.5cm
\caption[GMRT image and 21-cm absorption spectra of J0334+3921]{ Image of J0334+3921 (4C+39.12) with an rms noise of 0.1 mJy/beam.  The contour levels are
0.8$\times$($-1$, 1, 2, 4, 8, 16, 32, 64, 128, 256) mJy/beam.  
}
\label{J0334}
\end{figure}
%

%

\begin{figure}
\centerline{\vbox{
\psfig{figure=0415_map.ps,width=7.0cm,angle=-90}
\vskip -2.5cm
\hskip 0.2cm
\psfig{figure=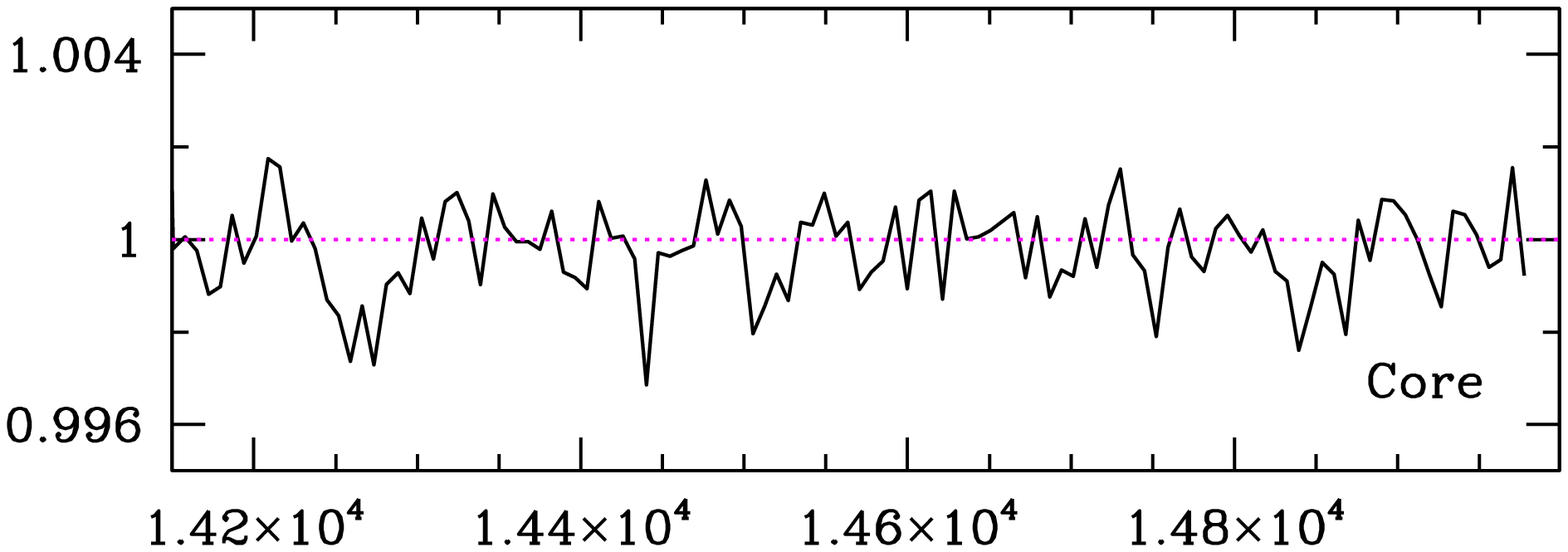,width=7.0cm,angle=0}
\vskip -4.0 cm
\hskip 0.2 cm
\psfig{figure=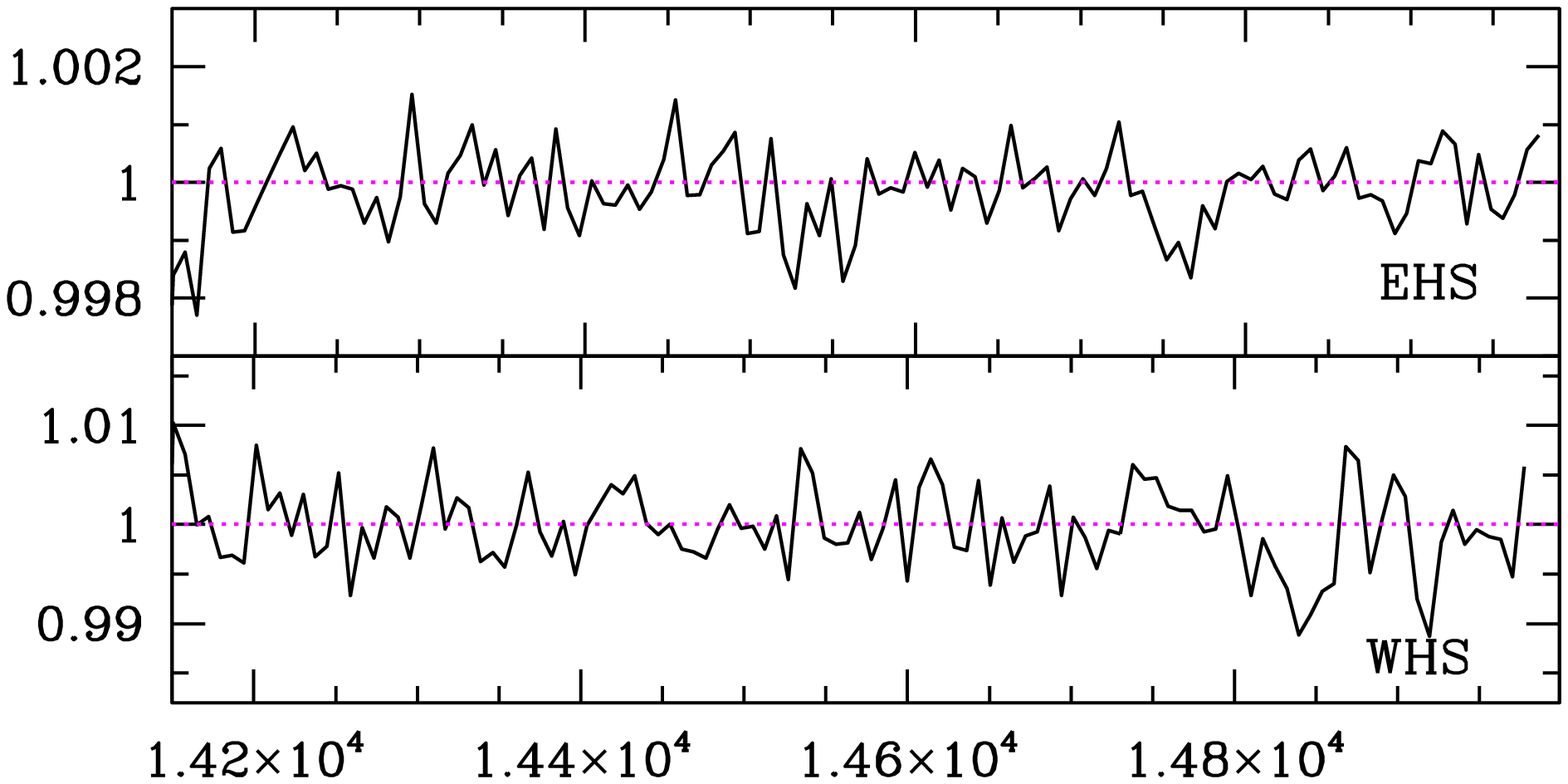,width=7.0cm,angle=0}
}}
\vskip -2.0 cm
\caption[GMRT image and 21-cm absorption spectra of J0418+3801]{ Image of J0418+3801 (3C\,111) with an rms noise of 0.6 mJy/beam.  The contour levels are
6$\times$($-1$, 1, 2, 4, 8, 16, 24, 32, 64, 128) mJy/beam.  
}
\label{J0415}
\end{figure}

%

\begin{figure}
\centerline{\vbox{
\psfig{figure=0652_MAP.PS,width=7.0cm,angle=0}
\vskip -2.8cm
\hskip +0.2cm
\psfig{figure=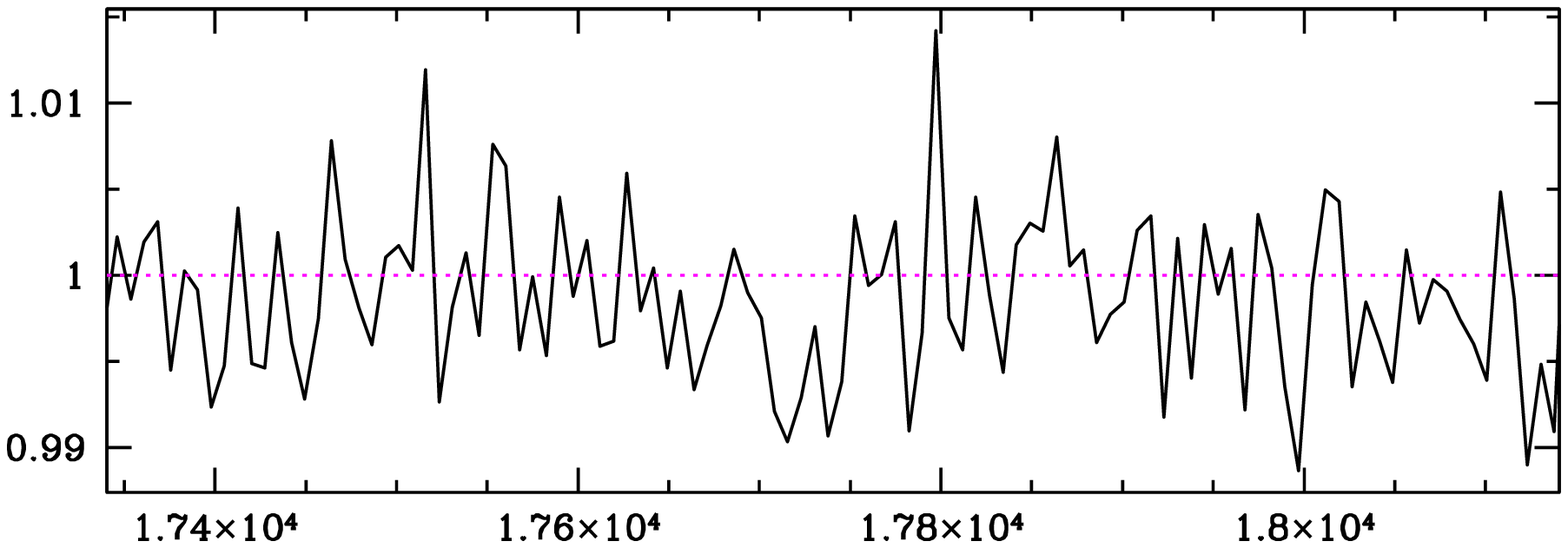,width=7.0cm,angle=0}
}}
\vskip -2.8 cm
\caption[GMRT image and 21-cm absorption spectra of J0656+4237]{ Image of J0656+4237 (4C\,42.22) with an 
rms noise of 0.24 mJy/beam.  The contour levels are
0.8$\times$($-1$, 1, 2, 4, 8, 16, 32, 64, 128) mJy/beam.  
}
\label{J0656}
\end{figure}

\begin{figure}
\centerline{\vbox{
\psfig{figure=0744_MAP.PS,width=7.0cm,angle=0}
\vskip -2.8cm
\hskip 0.2cm 
\psfig{figure=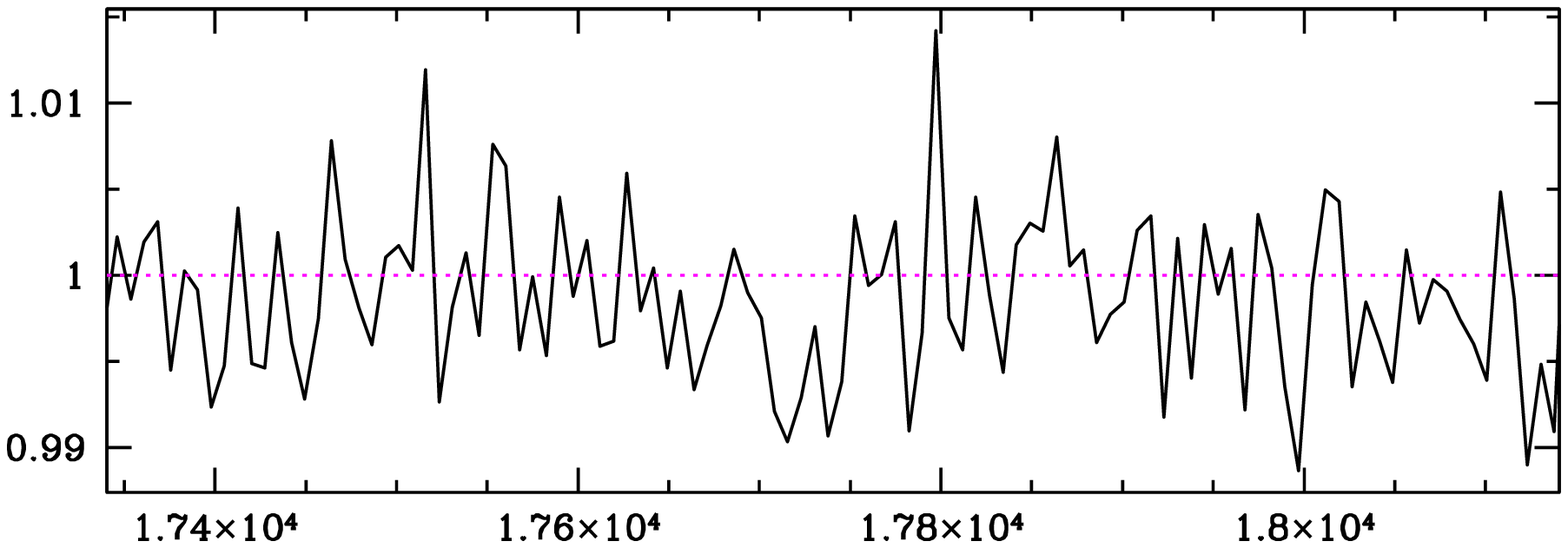,width=7.0cm,angle=0}
}}
\vskip -2.5 cm
\caption[GMRT image and 21-cm absorption spectra of J0748+5548]{ Image of J0748+5548 (DA\,240) with an rms noise of 0.37 mJy/beam.  The contour levels are
2.40$\times$($-1$, 1, 2, 4, 8, 16, 32, 64, 128, 256) mJy/beam.  
}
\label{J0748}
\end{figure}

%
%

\begin{figure}
\centerline{\vbox{
\psfig{figure=0755_MAP.PS,width=7.0cm,angle=-90}
\vskip -2.0cm
\hskip 0.2cm
\psfig{figure=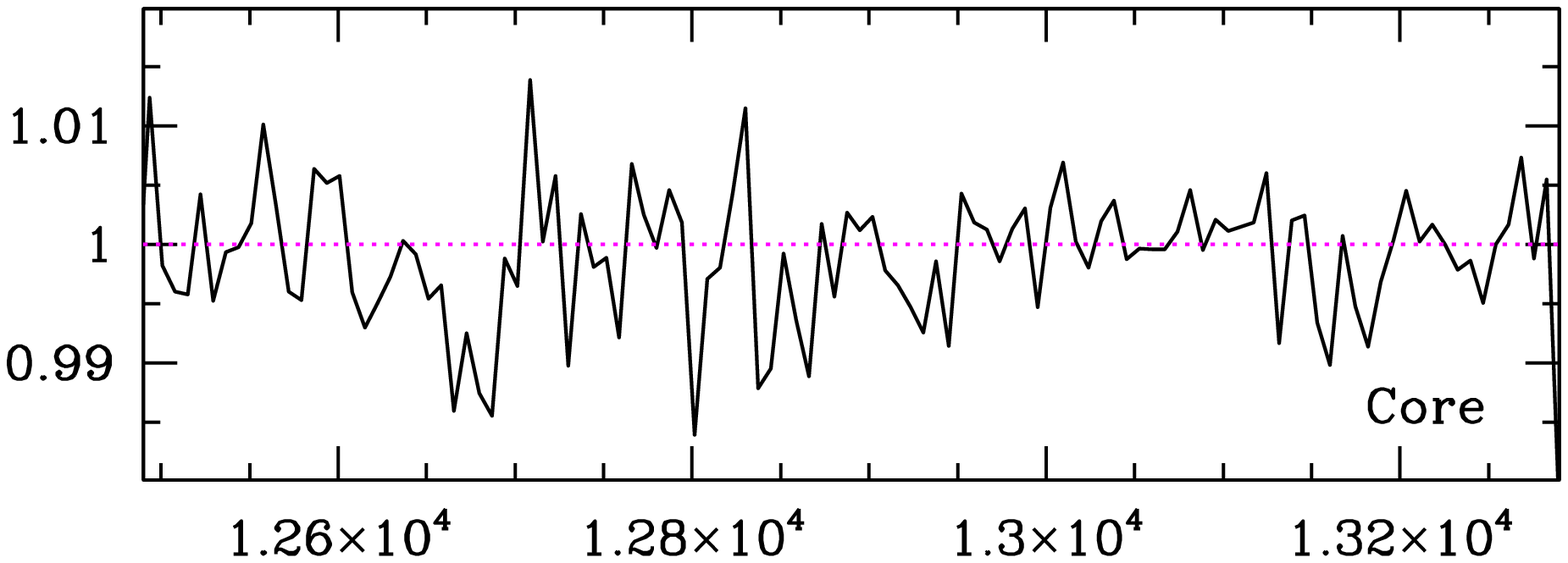,width=7.0cm,angle=0}
}}
\vskip -2.0cm
\caption[GMRT image and 21-cm absorption spectra of J0758+3747]{ Image of J0758+3747 (3C\,189) with an rms noise of 0.5 mJy/beam.  The contour levels are
2.0$\times$($-1$, 1, 2, 4, 8, 16, 32, 64, 128) mJy/beam. 
}
\label{J0755}
\end{figure}

\begin{figure}
\centerline{\vbox{
\psfig{figure=1144_MAP.PS,width=7.0cm,angle=-90}
\vskip -2.7cm
\hskip 0.2cm
\psfig{figure=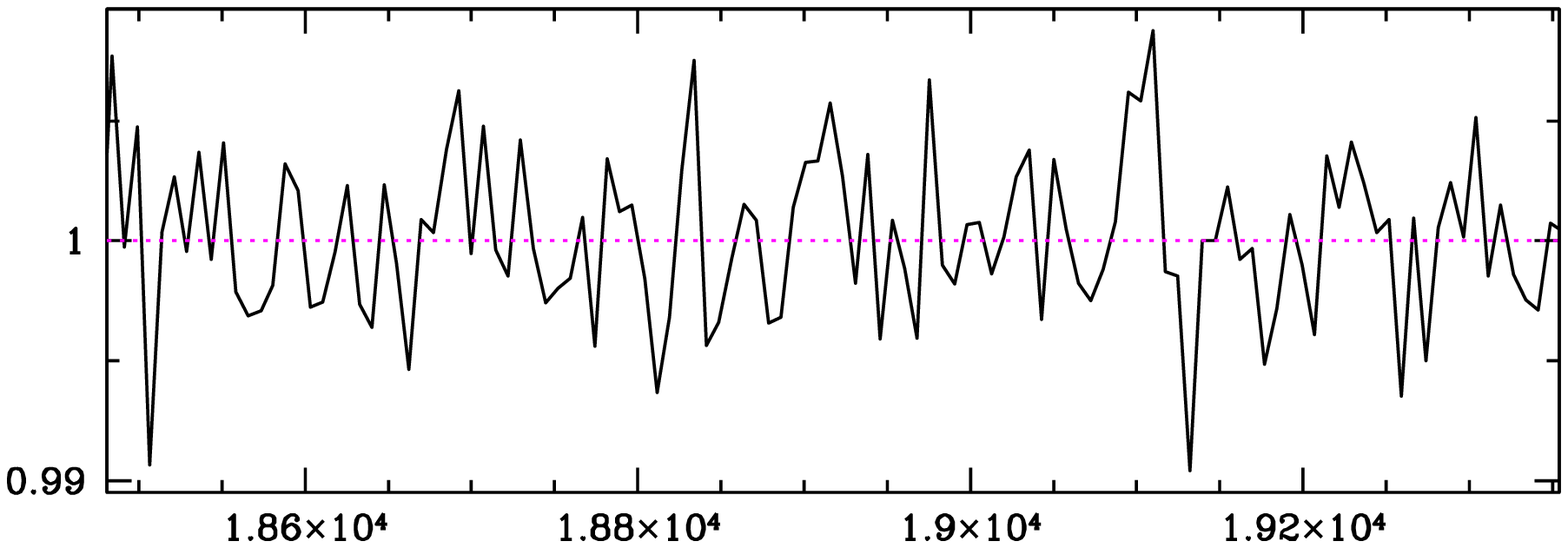,width=7.0cm,angle=0}
}}
\vskip -2.7cm
\caption[GMRT image and 21-cm absorption spectra of J1147+3501]{ Image of J1147+3501  with an rms noise of 0.54 mJy/beam.  The contour levels are
2.057$\times$($-1$, 1, 2, 4, 8, 16, 32, 64, 128, 256) mJy/beam. 
}
\label{J1147}
\end{figure}

\begin{figure}
\centerline{\vbox{
\psfig{figure=1626_MAP.PS,width=7.5cm,angle=-90}
\vskip -2.5cm
\hskip 0.2cm
\psfig{figure=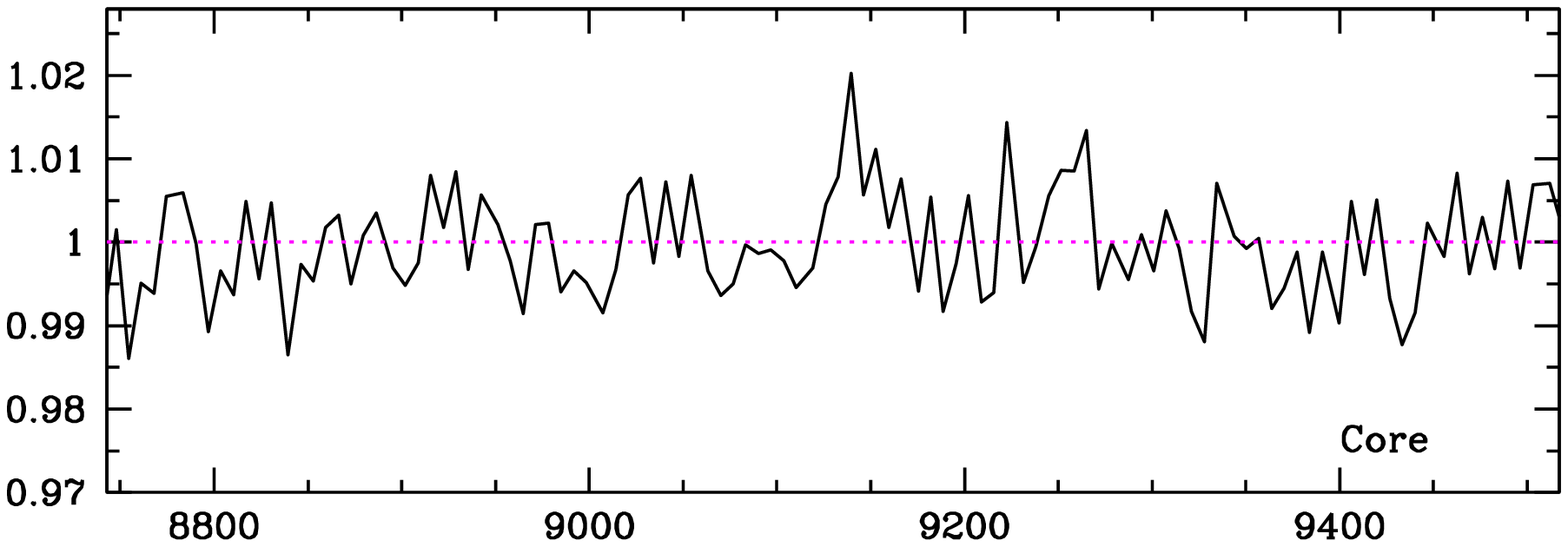,width=7.0cm,angle=0}
\vskip -3.6cm
\hskip 0.2cm
\psfig{figure=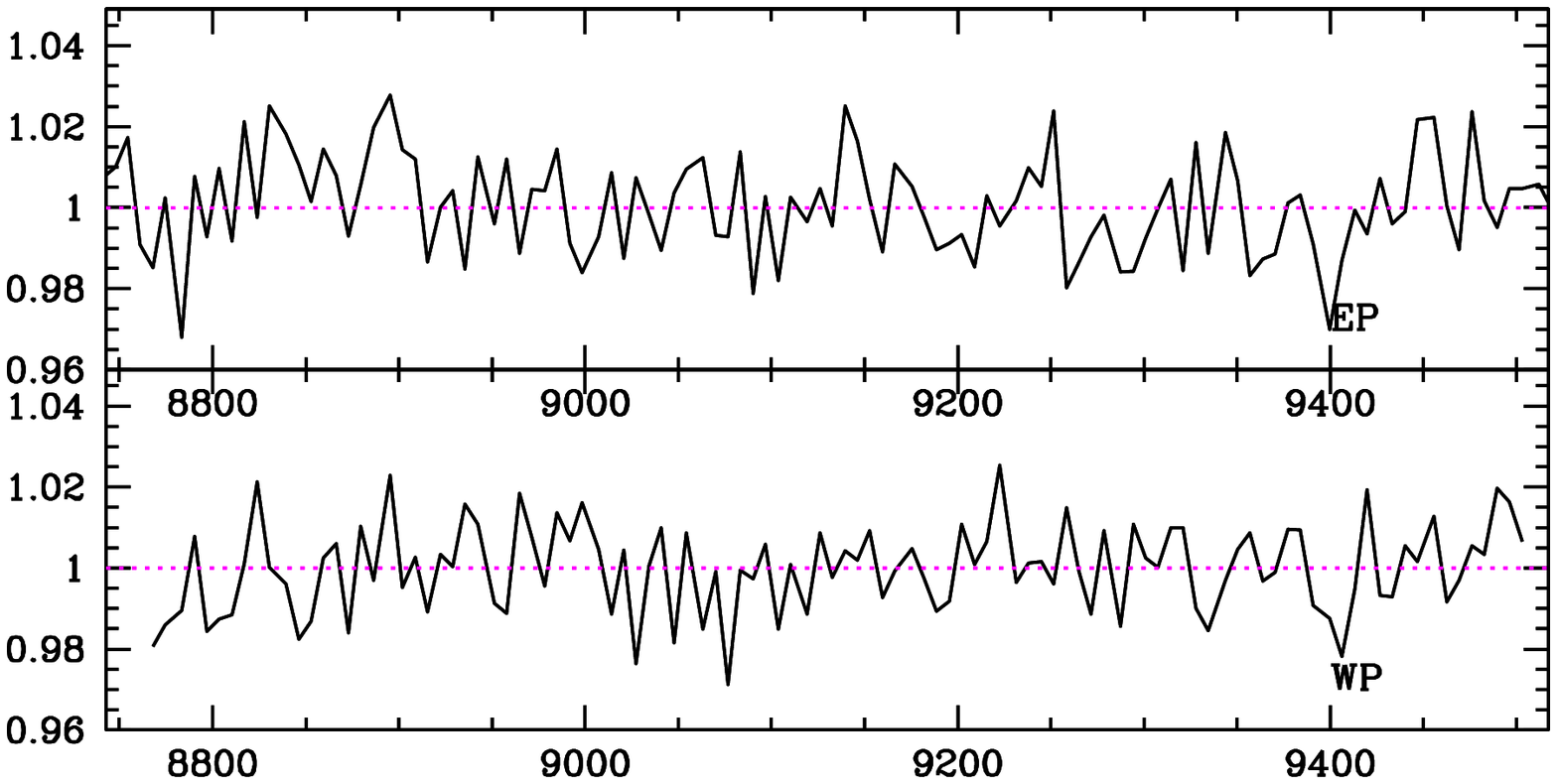,width=7.0cm,angle=0}
}}
\vskip -1.6cm
\caption[GMRT image and 21-cm absorption spectra of J1628+3933]{ Image of J1628+3933 with an rms noise of 0.46 mJy/beam. 
The contour levels are 5$\times$($-1$, 1, 2, 4, 8, 16, 32, 64, 128) mJy/beam.  
}
\label{J1628}
\end{figure}


\begin{figure}
\centerline{\vbox{
\psfig{figure=1744_MAP.PS,width=7.0cm,angle=-90}
\vskip -2.8cm
\hskip 0.2cm
\psfig{figure=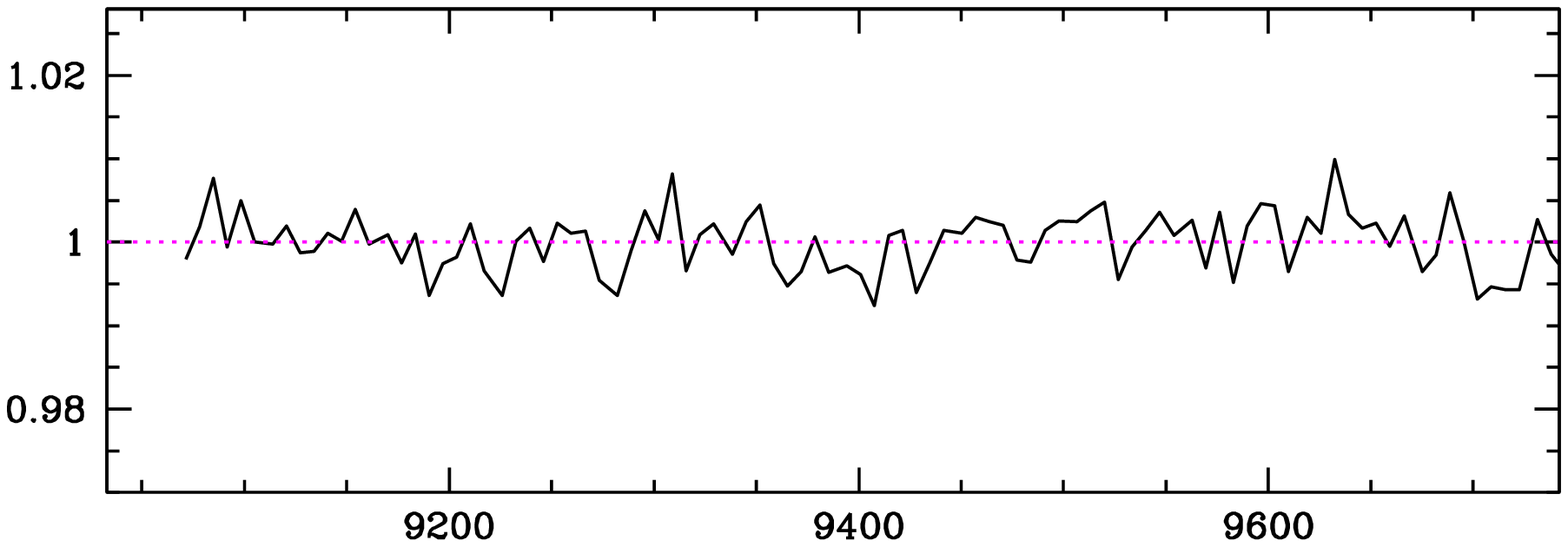,width=7.0cm,angle=0}
}}
\vskip-2.8cm
\caption[GMRT image and 21-cm absorption spectra of J1744+557]{ Image of J1744+557  with an rms noise of 0.48 mJy/beam.  The contour levels are
4$\times$($-1$, 1, 2, 4, 8, 16, 32, 64, 128, 256) mJy/beam.  
}
\label{J1744}
\end{figure}

\begin{figure}
\centerline{\vbox{
\vskip -0.5cm
\psfig{figure=1833_map.ps,width=7.0cm,angle=0}
\vskip -2.8cm
\hskip 0.2 cm
\psfig{figure=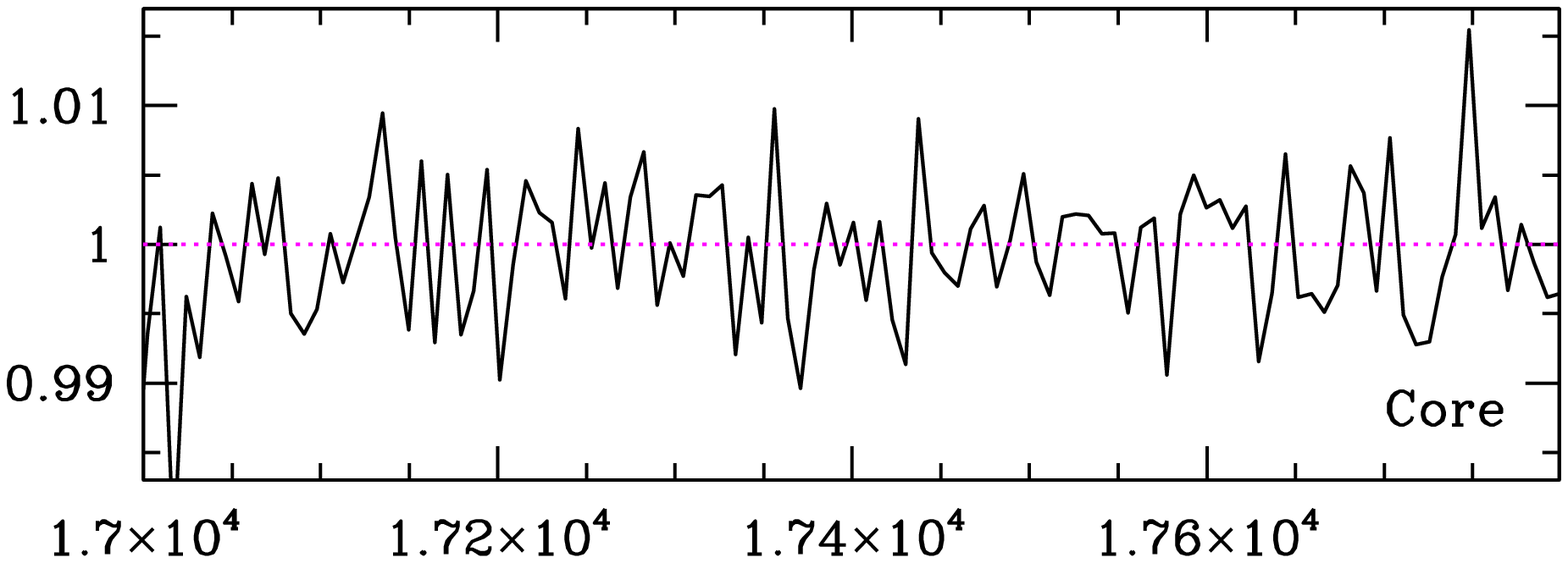,width=7.0cm,angle=0}
\vskip -3.7cm
\hskip 0.2cm
\psfig{figure=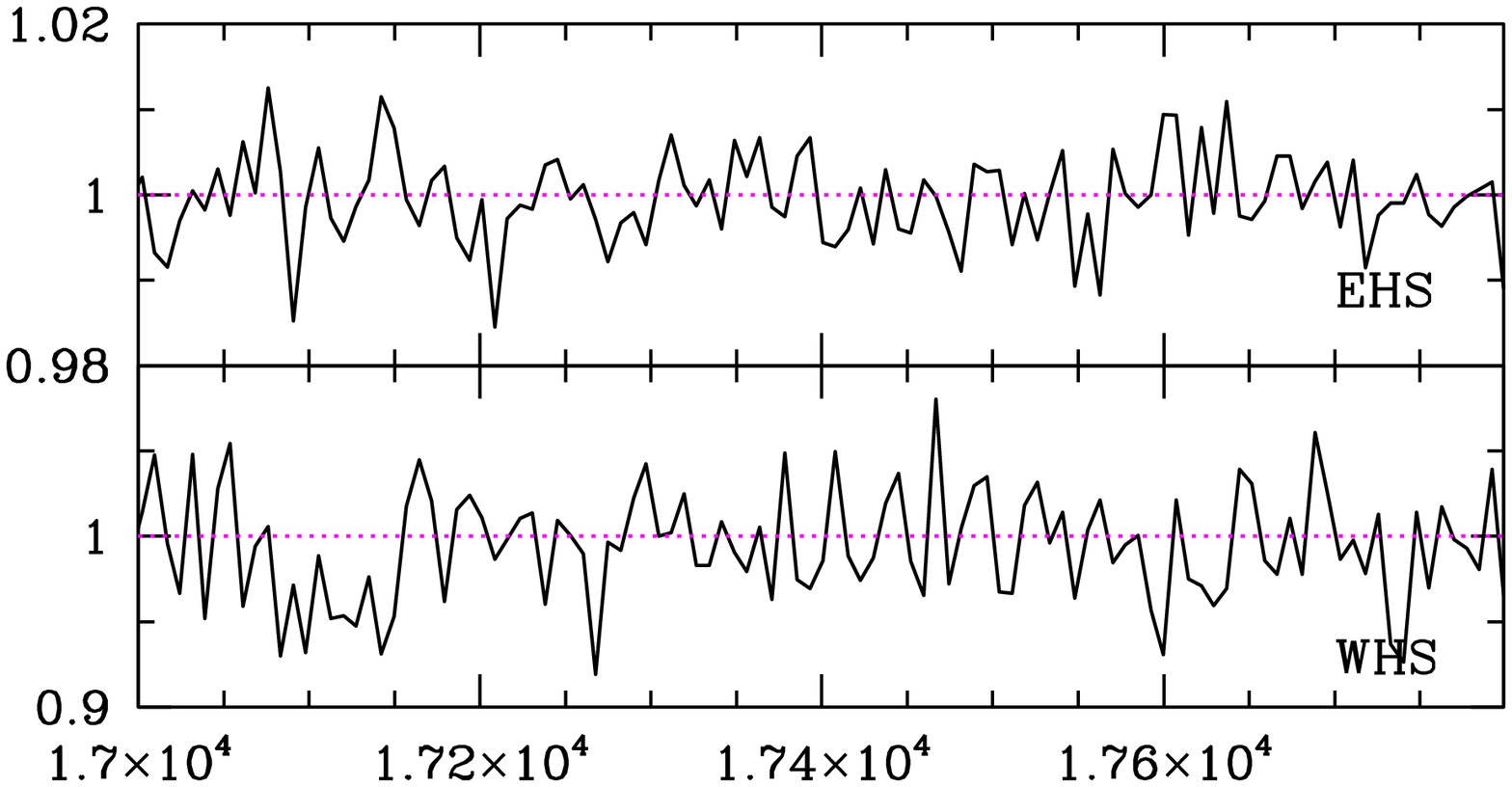,width=7.0cm,angle=0}
}}
\vskip -2.0cm
\caption[GMRT image and 21-cm absorption spectra of J1835+3241]{ Image of J1835+3241 (3C\,382) with an rms noise of 0.5 mJy/beam.  The contour levels are 2.35$\times$($-1$, 1, 2, 4, 8, 16, 32, 64, 128, 256) mJy/beam. 
}
\label{J1835}
\end{figure}
%
%

\begin{figure}
\centerline{\vbox{
\vskip -0.5cm
\psfig{figure=1834_MAP.PS,width=6.8cm,angle=0}
\vskip -2.8cm
\psfig{figure=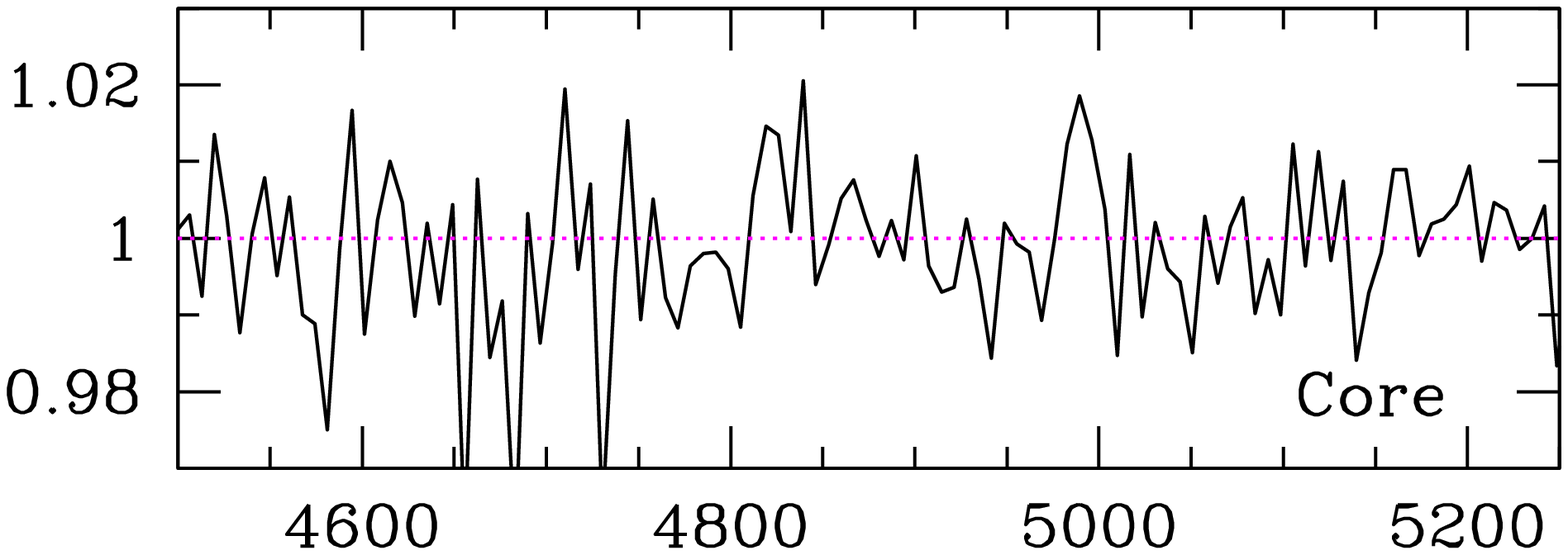,width=6.8cm,angle=0}
}}
\vskip -2.8cm
\hskip +0.0cm
\caption[GMRT image and 21-cm absorption spectra of J1836+1942]{ Image of J1836+1942 (PKS\,1834+196) with an rms noise of 0.8 mJy/beam.  The contour levels are
4.0$\times$($-1$, 1, 2, 4, 8, 16, 32, 64, 128, 256) mJy/beam.  
}
\label{J1836}
\end{figure}

\begin{figure}
\centerline{\vbox{
\vskip -1cm
\psfig{figure=1845_MAP.PS,width=6.8cm,angle=0}
\vskip -2.8cm
\hskip 0.2cm
\psfig{figure=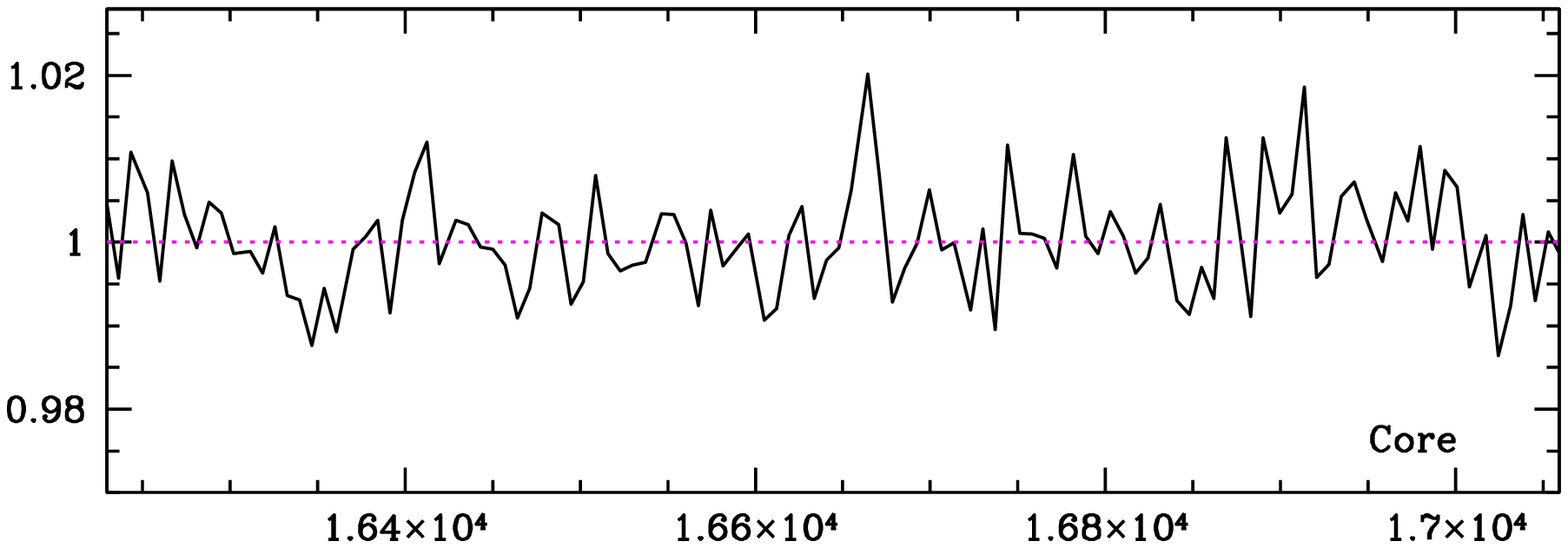,width=6.8cm,angle=0}
\vskip -4.0 cm
\hskip 0.2cm
\psfig{figure=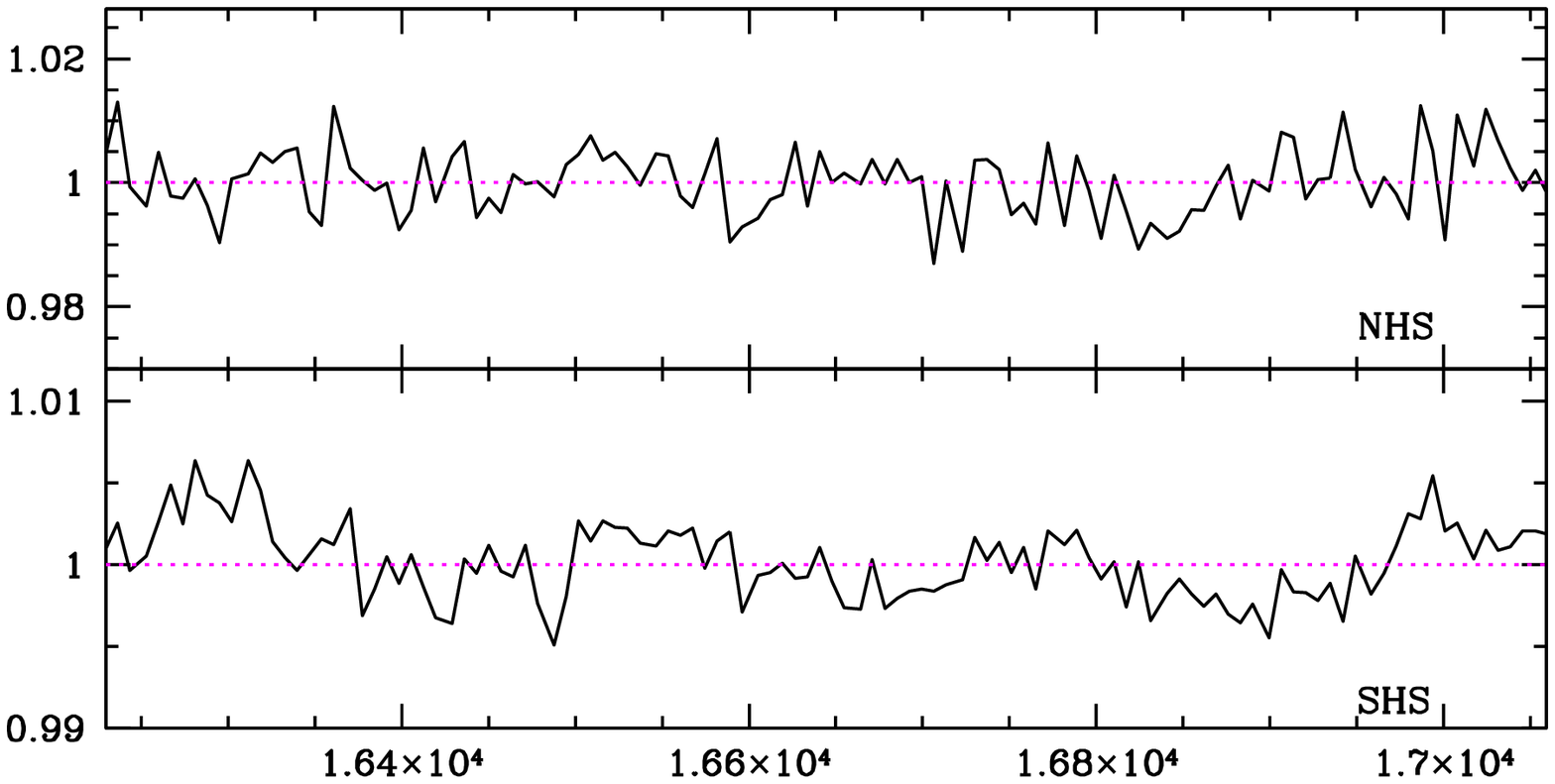,width=6.8cm,angle=0}
}}
\vskip -2.6cm
\hskip +0.0cm
\caption[GMRT image and 21-cm absorption spectra of J1842+7946]{ Image of J1842+7946 (3C390.3) with an rms noise of 0.59 mJy/beam.  The contour levels are
2.35$\times$($-1$, 1, 2, 4, 8, 16, 32, 64, 128, 256) mJy/beam.  
}
\label{J1842}
\end{figure}

\begin{figure}
\centerline{\vbox{
\psfig{figure=3C465_MAP.PS,width=6.8cm,angle=0}
\vskip -2.5cm
\hskip 0.2cm
\psfig{figure=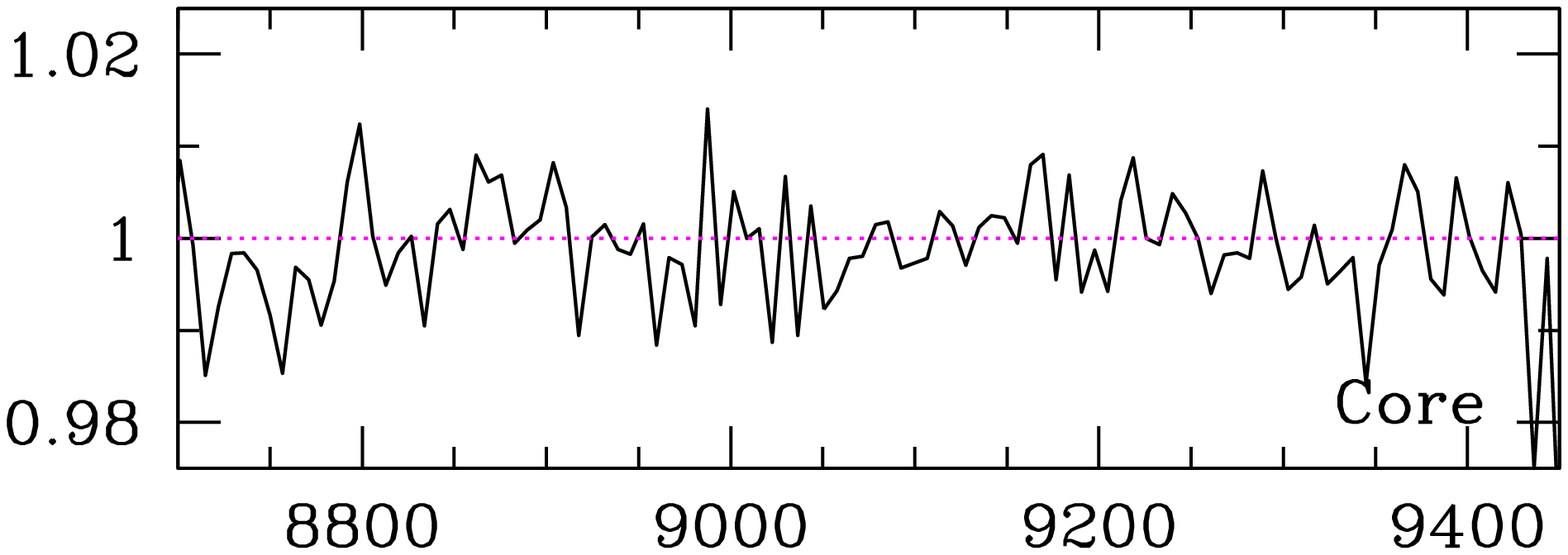,width=6.8cm,angle=0}
}}
\vskip -2.5cm
\caption[GMRT image and 21-cm absorption spectra of J2338+2701]
{Image of J2338+2701 (3C\,465) with an rms noise of 0.5 mJy/beam.  The contour levels are
2.5$\times$($-1$, 1, 2, 4, 8, 16, 32, 64, 128) mJy/beam.  
}
\label{j2338}
\end{figure}

\end{document}